\documentclass[lettersize,journal]{IEEEtran} 

\ifCLASSINFOpdf

\else

\fi
\usepackage[colorinlistoftodos]{todonotes}
\usepackage{amssymb}
\usepackage{amsmath,epsfig,amssymb,verbatim}
\usepackage{cite}
\usepackage[caption=false]{subfig}
\usepackage{array,algorithm,algorithmic}
\usepackage{amsmath,amsfonts,amssymb}
\usepackage{algorithmic}
\usepackage{algorithm}
\usepackage{array}
\usepackage{textcomp}
\usepackage{stfloats}
\usepackage{url}
\usepackage{verbatim}
\usepackage{cite}
\usepackage[normalem]{ulem}
\usepackage{mathtools}
\usepackage{comment}
\def\BibTeX{{\rm B\kern-.05em{\sc i\kern-.025em b}\kern-.08em
    T\kern-.1667em\lower.7ex\hbox{E}\kern-.125emX}}

\usepackage{tablefootnote}
\newtheorem{theorem}{Theorem}
\newtheorem{corollary}{Corollary}

\newtheorem{remark}{Remark}
\usepackage{enumerate}
\usepackage{multirow}
\usepackage{multicol}

\DeclareRobustCommand{\erase}{\bgroup\markoverwith{\textcolor{red}{\rule[.5ex]{2pt}{0.5pt}}}\ULon}

\def\BibTeX{{\rm B\kern-.05em{\sc i\kern-.025em b}\kern-.08em
    T\kern-.1667em\lower.7ex\hbox{E}\kern-.125emX}}
\usepackage{balance}

\begin{document}

\title{On the Coexistence  of OTFS Modulation with OFDM-based Communication Systems}
\author{Akram~Shafie,~\IEEEmembership{Member,~IEEE,}~Jinhong~Yuan,~\IEEEmembership{Fellow,~IEEE,} Paul Fitzpatrick,~\IEEEmembership{Senior~Member,~IEEE,} Taka Sakurai,~\IEEEmembership{Member,~IEEE,} and Yuting Fang,~\IEEEmembership{Member,~IEEE}
\thanks{A. Shafie and J. Yuan are with The University of New South Wales, Sydney, NSW, 2052, Australia (e-mail: akram.shafie@unsw.edu.au, j.yuan@unsw.edu.au).}
\thanks{P. Fitzpatrick, T. Sakurai, and Y. Fang are with Telstra Limited, Melbourne, Australia (e-mail: paul.g.fitzpatrick@team.telstra.com, taka.sakurai@team.telstra.com,~yuting.fang@team.telstra.com).}
\thanks{\textcolor{black}{The work was supported by the Australian Research Council (ARC) Linkage Project under Grant LP200301482.}}
\thanks{A preliminary version of this work was presented in 
2023 IEEE Global Communications (Globecom) Conference~\cite{akramGC2023_OTFSwithOFDM}.}
}

\maketitle
\begin{abstract}
We investigate the coexistence of orthogonal time-frequency space (OTFS) modulation with current fourth- and fifth-generation (4G/5G) communication systems that primarily use orthogonal frequency-division multiplexing (OFDM) waveforms. We first derive the input-output-relation of OTFS in the considered coexisting system. In this derivation, we consider (i) the inclusion of multiple cyclic prefixes (CPs) with unequal lengths to the OTFS signal and (ii) edge carrier unloading (ECU), to account for the impacts of CP length, frame structure, and subcarrier arrangement described in 3GPP standards for 4G/5G systems. Our analysis reveals that the inclusion of multiple CPs to the OTFS signal and ECU lead to the channel response exhibiting spreading effects/leakage along the Doppler and delay dimensions, respectively. Consequently, the effective sampled delay-Doppler (DD) domain channel model for OTFS in coexisting systems may exhibit reduced sparsity. We also show that the effective DD domain channel coefficients for OTFS in coexisting systems are influenced by the unequal lengths of CPs. Subsequently, we propose an interference cancellation-based channel estimation (CE) technique for OTFS in coexisting systems.  Through numerical results, we validate our analysis, highlight the importance of not ignoring the unequal lengths of CPs during signal detection, and show the significance of the proposed CE technique.
\end{abstract}


\section{Introduction}

\IEEEPARstart{O}{rthogonal}
time-frequency space (OTFS) modulation has emerged as a promising contender  to achieve reliable communications in high-mobility scenarios, particularly in the context of the fifth-generation (5G) and beyond eras~\cite{2021_WCM_JH_OTFS}. In contrast to conventional modulation schemes operating in the time and/or frequency domain, OTFS is a two-dimensional (2D) modulation scheme that carries the to-be-transmitted information over the delay-Doppler (DD) domain~\cite{2022_TWC_Lorenzo_OTFSvsOFDMcomparison}.
Due to its inherent capabilities in effectively addressing the challenges posed by the Doppler effect, OTFS has garnered significant interest from both industry and academia since its introduction by Hadani \textit{et al.} in 2017~\cite{2017_WCNC_OTFS_Haddani}.

\begin{figure}[t]
\centering
\vspace{-2mm} 
\includegraphics[width=0.99\columnwidth]{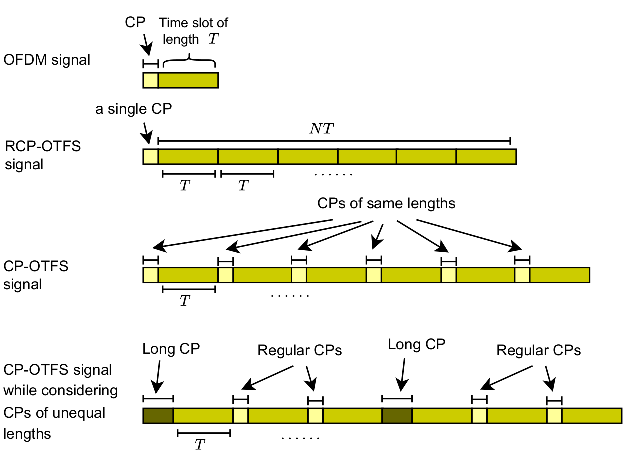}
\vspace{-2mm}
\caption{\textcolor{black}{A simplified illustration of the time domain representation of the transmitted signals of OFDM, RCP-OTFS, CP-OTFS, and CP-OTFS while considering CPs of unequal lengths.}}\label{Fig:AllSignals} 
\end{figure}

One variant of OTFS, which is commonly referred to as reduced cyclic prefix (CP)-based OTFS (RCP-OTFS), has been widely explored in the literature~\cite{2022_EmanuelBook_DDCom,2020_TVT_Viterbo_RakeDFEforOTFS,2018_TWC_Viterbo_OTFS_InterferenceCancellation,2000_TCOM_VOFDM}. This is due to its ability to easily couple with the sparse DD domain representation of doubly-selective channels.
In RCP-OTFS, as shown in  Fig.~\ref{Fig:AllSignals}, a single CP  is included in the entire OTFS signal (i.e., a single CP  for multiple time slots with a total duration $NT$). 
Despite its advantages,  there are several unresolved challenges associated with the practical implementation of RCP-OTFS, particularly concerning its transceiver design~\cite{2022_ComLet_Cheng_OTFSErrorPerformance,2019_TCT_ReducedCPOTFS,2022_TWC_JH_ODDM}. Transceiver designs that employ  rectangular pulses are commonly adopted in the literature for RCP-OTFS~\cite{2020_TVT_Viterbo_RakeDFEforOTFS,2018_TWC_Viterbo_OTFS_InterferenceCancellation,2019_TCT_ReducedCPOTFS}; however,  their practical adoption might prove challenging as they lead to high out-of-band emission (OOBE)~\cite{2019_TCT_ReducedCPOTFS,2022_ComLet_Cheng_OTFSErrorPerformance}.
To address the OOBE, on the one hand,~\cite{2019_TCT_ReducedCPOTFS} proposed a transceiver design based on pulse shaping. On the other hand,~\cite{2022_TWC_JH_ODDM} proposed a practically implementable transceiver design based on DD domain multi-carrier (DDMC) modulation.
Although these proposed transceiver designs successfully mitigate the OOBE, a significant limitation arises from the fact that they differ from those employed in current fourth-generation (4G) and 5G communication systems.\footnote{We note that any new modulation scheme or variant is not obliged to smoothly coexist with current 4G and 5G systems.
Nonetheless,  in order to protect the substantial investments made in the transceivers of current 4G and 5G systems from becoming
obsolete, commercial vendors may favor new modulation schemes or variants that can smoothly coexist with current 4G and 5G systems, even if their performance is comparatively lower~\cite{5GCostAustralia}. 
} 

\textcolor{black}{
Particularly, the current 4G and 5G communication systems primarily employ orthogonal frequency-division multiplexing (OFDM) waveforms~\cite{4GwithOFDM,2018_ComStandardMag_OFDMNumerology}. The OFDM signals utilize a CP  for every OFDM symbol duration or time slot $T$ (see Fig.~\ref{Fig:AllSignals}). Differently, the transceiver designs proposed in~\cite{2019_TCT_ReducedCPOTFS,2022_TWC_JH_ODDM} utilize a single CP for multiple time slots with total duration $NT$ (see Fig.~\ref{Fig:AllSignals}).
Additionally, as will be discussed later, current 4G and 5G communication systems employ CPs of unequal/different lengths and edge carrier unloading
(ECU)~\cite{3GPPStandOFDM,5GOFDMWebpage,2018_ComStandardMag_OFDMNumerology,2011_TWC_OFDMECURef1,2009_TWC_OFDMECURef2,2016_ComLet_OFDMECURef3}. However, the transceiver designs proposed in~\cite{2019_TCT_ReducedCPOTFS,2022_TWC_JH_ODDM} considered neither CPs of unequal lengths nor ECU. These differences make it impossible for RCP-OTFS to coexist with  (a.k.a. to be implemented on top of or to be backward compatible with) current 4G and 5G communication systems.
This has prompted the exploration of alternative OTFS variants that can yield transceiver designs ensuring both minimal OOBE and 
smooth coexistence with current 4G and 5G communication systems.}

One such variant, known as CP-based OTFS (CP-OTFS), serves as an alternative to RCP-OTFS~\cite{2022_EmanuelBook_DDCom}.  It differs  from RCP-OTFS in terms of how CPs are included in the OTFS signal.
Specifically, in CP-OTFS, multiple CPs are included in the OTFS signal \textcolor{black}{(see Fig.~\ref{Fig:AllSignals})}. 
Due to the fact that current 4G and 5G communication systems primarily employ OFDM waveforms while utilizing different methods to eliminate OOBE and the fact that CP-OTFS can be viewed as precoded-OFDM~\cite{2018_PIMRC_IterativeDetectionforDSChannel}, or a collection of multiple OFDM signals, CP-OTFS emerges as a promising candidate for ensuring both OOBE minimization and seamless coexistence with current 4G and 5G communication systems.
This serves as our motivation to investigate CP-OTFS while focusing on its coexistence with OFDM systems.

Several studies have delved into the investigation of CP-OTFS~\cite{2018_WCL_ModemStructureforOFDMBasedOTFS,2018_ICC_OTFSwithOFDM,2022_IEEEAccess_Viterbo_GeneralIOR,2019_ICC_CPOTFSwithMassiveMIMO}.  The CP-OTFS was first  investigated in~\cite{2018_WCL_ModemStructureforOFDMBasedOTFS}, where its DD domain channel matrix was derived. Moreover,~\cite{2018_ICC_OTFSwithOFDM} proposed a modem structure for CP-OTFS implementation. Furthermore,~\cite{2022_IEEEAccess_Viterbo_GeneralIOR} derived the delay-time domain input-output relation (IOR) for CP-OTFS. 
Additionally,~\cite{2019_ICC_CPOTFSwithMassiveMIMO} investigated CP-OTFS in massive multiple-input multiple-output systems. 
Despite the progress, we note that the focus of~\cite{2018_ICC_OTFSwithOFDM,2018_WCL_ModemStructureforOFDMBasedOTFS,2022_IEEEAccess_Viterbo_GeneralIOR,2019_ICC_CPOTFSwithMassiveMIMO} was not on comprehending the coexistence of OTFS with OFDM systems.
Thus, the impact of crucial aspects of OFDM systems, which could significantly influence the performance of OTFS in coexistence scenarios, was not thoroughly considered in~\cite{2018_ICC_OTFSwithOFDM,2018_WCL_ModemStructureforOFDMBasedOTFS,2022_IEEEAccess_Viterbo_GeneralIOR,2019_ICC_CPOTFSwithMassiveMIMO}.

When OTFS coexists with OFDM systems, first, it is necessary to account for the impacts of CP and frame structures specified in the 3GPP standards for OFDM systems~\cite{3GPPStandOFDM,5GOFDMWebpage,2018_ComStandardMag_OFDMNumerology}. According to 3GPP standards, the downlink and uplink transmissions are arranged into \textit{frames}, which are then divided into \textit{sub-frames}. Thereafter, \textit{sub-frames} are further divided into two equally-sized half-frames of 0.5 ms that contain multiple OFDM signals, which we refer to as  \textit{time windows}. 
To obtain reasonable time diversity, the utilization of multiple time windows may become necessary to transmit an OTFS signal using OFDM systems. Additionally,  within each time window, the first OFDM signal is assigned a CP of longer length than the subsequent OFDM signals~\cite{3GPPStandOFDM,5GOFDMWebpage,2018_ComStandardMag_OFDMNumerology}. Consequently, it becomes imperative to investigate CP-OTFS in coexisting systems while considering unequal lengths of CPs \textcolor{black}{(see Fig.~\ref{Fig:AllSignals})}.\footnote{\textcolor{black}{We clarify that when we refer to "OTFS modulation with CPs of unequal lengths" in this work, we are not implying that each CP in the OTFS signal has a distinct length from each other.  Instead, our consideration is that not all the CPs in the OTFS signal have the same length, although some CPs in the OTFS signal may share the same length.}}

Another crucial aspect to consider when OTFS coexists with OFDM systems is the impact of ECU/null subcarriers~\cite{3GPPStandOFDM,2011_TWC_OFDMECURef1,2009_TWC_OFDMECURef2,2016_ComLet_OFDMECURef3}.
The ECU refers to 
disallowing the transmission of symbol/energy on the edge subcarriers in an OFDM signal.
This technique is widely adopted in practical OFDM systems 
for two purposes~\cite{2023_ODDM_TCOM}. Firstly, ECU enables the mitigation of OOBE, thereby ensuring that the transmitted OFDM signals meet the spectral mask constraints. Secondly, ECU provides the flexibility to set the total number of subcarriers in an OFDM signal to a power of two, thereby facilitating the easier
hardware implementation of OFDM transceivers~\cite{2011_TWC_OFDMECURef1,2009_TWC_OFDMECURef2}. 
This motivates the investigation of OTFS in coexisting systems while considering ECU. 

Despite the significance of both CPs of unequal lengths and ECU, their impacts have not been thoroughly explored in the existing literature. Very recently,~\cite{2020_IEEEAccess_IITKhar_CPOTFS_CEandSynchronizationErrors,2022_IEEEWCL_IIRBhub_CarrierFreuqencyOffsetDCOffsetIQImbalance} explored the coexistence of OTFS with OFDM systems. These studies focused on characterizing the impairments arising from the carrier frequency, frame timing, and direct current offsets.  
However, they did not investigate the impact of either CPs of unequal lengths or ECU.

In this work, we investigate the coexistence of OTFS modulation with current 4G/5G communication systems that use OFDM waveforms. 
The main contributions of this work are fourfold, and are summarized as follows:


\textbf{1)} We first derive the element-wise IOR of OTFS in a coexisting system that uses OFDM waveforms. In this derivation, firstly, we consider  
the inclusion of multiple CPs with unequal lengths to the OTFS signal, 
to account for the impacts of CP and frame structures described in the 3GPP standards. Secondly, we consider the impact of ECU  which is widely adopted in practical OFDM systems to mitigate  OOBE.

\textbf{2)} Through analytical deliberation, we reveal several important findings, which are listed as follows:
\begin{itemize}
  \item  \textcolor{black}{Even when the Dopplers and delays of all the propagation paths are on-grid (a.k.a integer Dopplers and delays~\cite{2018_TWC_Viterbo_OTFS_InterferenceCancellation,2022_EmanuelBook_DDCom,2020_TVT_Viterbo_RakeDFEforOTFS}), the inclusion of multiple CPs to the OTFS signal leads to every single propagation path being perceived as multiple taps with the same delay and different Dopplers, i.e.,  channel response exhibits spreading effects/leakage along the Doppler dimension;}
  \item  Even when the Dopplers and delays of all the propagation paths are on-grid, ECU  leads to the channel response exhibiting spreading effects along the delay dimension;\footnote{\textcolor{black}{For OTFS in coexisting systems with off-grid Dopplers and off-grid delays  (a.k.a fractional Dopplers and delays~\cite{2018_TWC_Viterbo_OTFS_InterferenceCancellation,2022_EmanuelBook_DDCom,2020_TVT_Viterbo_RakeDFEforOTFS}), firstly, the combination of off-grid Doppler and the inclusion of multiple CPs to the OTFS signal causes the channel response to exhibit spreading effects along the Doppler dimension. Secondly, the combination of off-grid delay and ECU causes the channel response to exhibit spreading effects along the delay dimension.}}
  \item As a result of this spreading, \textit{the effective sampled DD domain channel model for OTFS in coexisting systems may exhibit reduced sparsity};
  \item The effective DD domain channel coefficients for OTFS in coexisting systems are influenced by the unequal lengths of the CPs. 
\end{itemize}

\textbf{3)} We next propose an embedded pilot-aided channel estimation (CE) technique for OTFS in coexisting systems.
In scenarios where channel response exhibits spreading effects along the Doppler (or/and delay) dimension, such as OTFS in coexisting systems which is the focus of this work, the state-of-the-art threshold-based CE technique proposed in~\cite{2019_TVT_YiHong_ChannelEstimationforOTFS} may demand extremely high pilot power to accurately characterize the channel. This can cause high peak-to-average power ratio (PAPR) during the practical implementation of OTFS in coexisting systems~\cite{2019_TVT_YiHong_ChannelEstimationforOTFS,2022_TWC_PAPRofOTFS,2022_TWC_Zhigiang_OffGridCEforOTFS}.  To tackle this challenge, 
we propose an interference cancellation-based CE technique that estimates the channel coefficient and the Doppler of each propagation path, using the observed 
effective DD domain channel coefficients corresponding to the propagation path.

\textbf{4)} We perform extensive numerical analysis to attain novel insights on our considerations. We first verify our analysis by comparing it with simulation
results. We then show that ignoring the impact of unequal lengths of CPs during signal detection can degrade the bit error rate (BER) of OTFS in coexisting systems.
We also show that the proposed CE technique outperforms the threshold-based CE technique and the BER of the proposed CE technique approaches that achieved with perfect channel state information~(CSI).

\section{System model}

\begin{figure*}[t]
\hspace{-8mm}
\includegraphics[width=2.25\columnwidth]{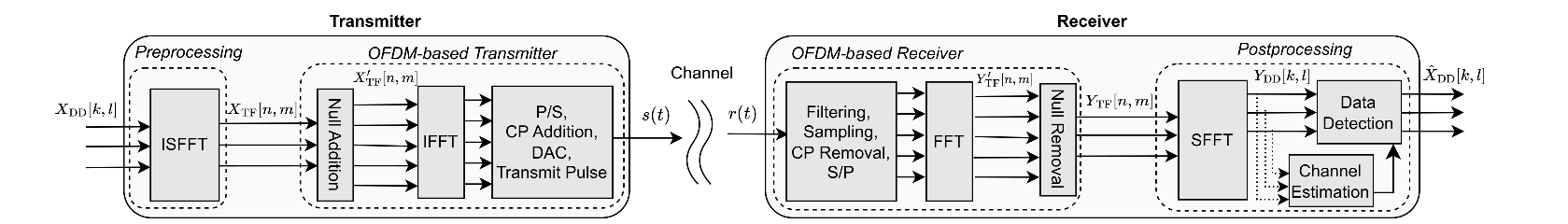}
\caption{Illustration of the system model considered in this work, where OTFS modulation coexists with the current OFDM-based communication system while introducing preprocessing at the transmitter and postprocessing at the receiver.}\label{Fig:SysModel} 
\end{figure*}

We focus on a communication scenario where  OTFS modulation coexists with current 4G/5G communication systems that use OFDM waveforms, while introducing preprocessing at the transmitter and postprocessing at the receiver. The schematic representation of the considered system is shown in Fig.~\ref{Fig:SysModel}.

\subsection{Transmitter}

Consider  OTFS modulation in which $NM$ message symbols are mapped in the DD domain, where $N$ and $M$ denote the number of Doppler and delay elements within the OTFS signal, respectively. 
We denote the $(k,l)$-th element of the DD domain signal as $X_{\textrm{DD}}[k,l]$, where $k$ and $l$ denote the Doppler and delay indices, respectively, with $k\in\mathcal{N}={\{0,1,..., N-1\}}$ and $l\in\mathcal{M}={\{0,1,..., M-1\}}$. 

At the preprocessing stage of the transmitter, a time-frequency (TF) domain signal 
is obtained by applying inverse simplectic finite Fourier transform (ISFFT) on $X_{\textrm{DD}}$~\cite{2017_WCNC_OTFS_Haddani}. Specifically, the $(n,m)$-th symbol of the TF domain signal, $X_{\textrm{TF}}[n,m]$, is obtained as
\begin{align}\label{Equ:Xtf}
X_{\textrm{TF}}[n,m] &= \frac{1}{\sqrt{NM}}\sum_{k=0}^{N-1} \sum_{l=0}^{M-1} X_{\textrm{DD}}[k,l]e^{j2\pi(\frac{nk}{N}-\frac{ml}{M})},
\end{align}
where $n\in\mathcal{N},m\in\mathcal{M}$, and 
$n$ and $m$ denote the time and frequency indices, respectively.

\begin{figure*}[t]
\centering
\vspace{-2mm} 
\includegraphics[width=1.8\columnwidth]{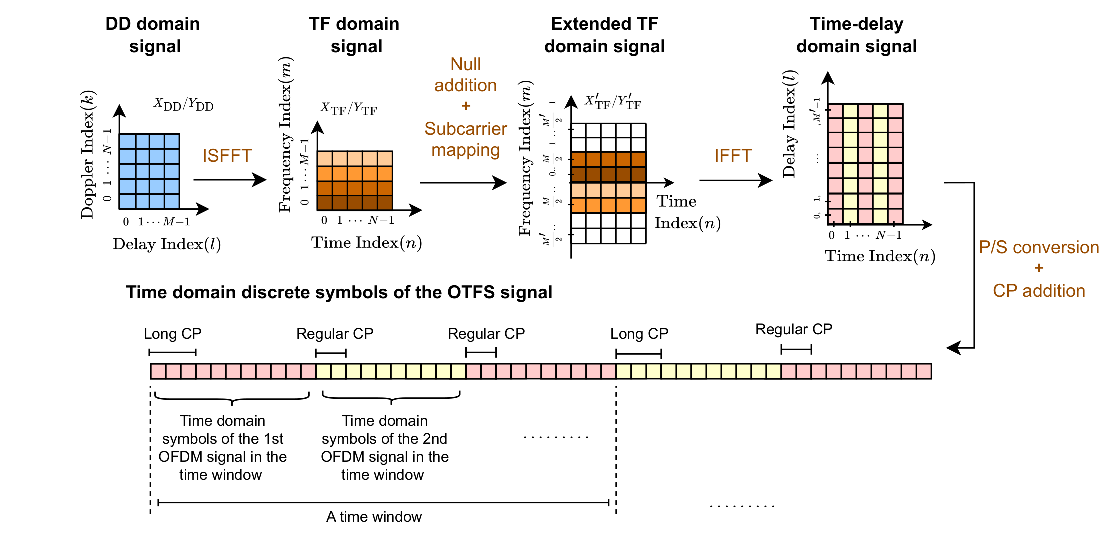}
\vspace{-2mm}
\caption{\textcolor{black}{Illustration of the DD domain, TF domain, extended TF domain, time-delay domain, and the time domain symbols of the OTFS signal at the transmitter. In this figure, we consider $N=5$, $M=4$, $M'=8$, $S=3$, $T_{\mathrm{long}}^{\mathrm{cp}}=3\times \frac{T}{M'}$, and $T_{\mathrm{reg}}^{\mathrm{cp}}=2\times \frac{T}{M'}$.}}\label{Fig:SigTrans} 
\end{figure*}

At the OFDM-based transmitter, $N$ OFDM signals are generated
using $X_{\textrm{TF}}$. First, an extended TF domain signal is obtained using null symbols and $X_{\textrm{TF}}$. 
Specifically, the $(n,m)$-th symbol of the extended TF signal, $X'_{\textrm{TF}}[n,m]$, 
is expressed as 
\begin{align}
\!\!\!&X'_{\textrm{TF}}[n,m] {=} \begin{cases}
0, & -\frac{M'}{2} \leqslant m < -\frac{M}{2} ,\\
X_{\textrm{TF}}[n,m+M], & -\frac{M}{2} \leqslant m < 0 ,\\
X_{\textrm{TF}}[n,m], & 0 \leqslant m < \frac{M}{2} ,\\
0, & \frac{M}{2} \leqslant m \leqslant \frac{M'}{2}-1,\!\!\!
\end{cases} 
 \label{Equ:Xtfprim}
\end{align}
where $n\in\mathcal{N},m\in\mathcal{M}'$, $M'$ denotes the total number of sub-carriers in the OFDM signal, and $\mathcal{M}'\in{\{{-}\frac{M'}{2},{-}\frac{M'}{2}{+}1,\cdots,{-}1,0,1,\cdots \frac{M'}{2}{-}1\}}$~\cite{2023_ODDM_TCOM}.
We clarify that practical OFDM systems adopt edge carrier unloading (ECU)/null subcarriers, which refers to 
disallowing the transmission of symbol/energy on the edge subcarriers in an OFDM signal.
The consideration of ECU in OFDM systems serves two main purposes~\cite{2023_ODDM_TCOM}.  Firstly, it effectively mitigates OOBE, thereby ensuring that the transmitted OFDM signals meet the spectral mask constraints~\cite{2011_TWC_OFDMECURef1,2009_TWC_OFDMECURef2}. Secondly, ECU provides the flexibility to set the total number of subcarriers in an OFDM signal to a power of two. This, as will be discussed in Remark~\ref{Rem:SelectionofMMprim}, facilitates the easier hardware implementation of OFDM transceivers. This utilization of ECU is reflected in $X'_{\textrm{TF}}$ with the inclusion of null symbols, and further illustrated in Fig.~\ref{Fig:SigTrans}.

Next, symbols in $X'_{\textrm{TF}}$ are sent through an inverse fast Fourier transform (IFFT) operation and a parallel-to-serial (P/S) converter to obtain the time domain discrete symbols. Finally, a cyclically extended version of the time domain discrete symbols are sent through a digital-to-analog converter (DAC), which is followed by a transmit pulse or time domain windowing operation. Considering the aforementioned operations at the OFDM-based transmitter, the $n$th OFDM signal associated with the OTFS signal, $x_n(t)$, is obtained as~\cite{2023_ODDM_TCOM} 
\begin{align}\label{st2} 
x_n(t)&=\frac{1}{\sqrt{M'}}\!\!\!\sum _{m=-\frac{M'}{2}}^{\frac{M'}{2}-1}X'_{\textrm{TF}}[n,m]e^{j2\pi m \Delta {f} t} g_n(t)
,~~\forall~ n\in\mathcal{N},
\end{align}
where $\Delta f$ denotes the subcarrier spacing and $g_n(t)$ denotes the transmitter pulse of the $n$th OFDM signal, given by 
\begin{align}\label{Equ:gtx}
g_n(t)&=\begin{cases}
1, & ~~~~~-T^{\mathrm{cp}}_n\leqslant t\leqslant T,\\
0, & ~~~~~~\textrm{elsewhere},
\end{cases}~~~~~~\forall~ n\in\mathcal{N},
\end{align}
where $T^{\mathrm{cp}}_n$ denotes the CP length of the $n$th OFDM signal. 
In order to appropriately fulfill the orthogonality conditions for OFDM signals,  $T$ and $\Delta f$ are let to adhere to the relationship $T=\frac{1}{\Delta f}$~\cite{3GPPStandOFDM}.
As can be observed in~\eqref{st2}, and further illustrated in Fig.~\ref{Fig:SigTrans}, the highest frequency to which the symbols of $X'_{\textrm{TF}}[n,m]$ are modulated is limited to $\frac{M'}{2T}$. This is to ensure that $x_n(t)$ is strictly bandlimited
to $[-\frac{M'}{2T},\frac{M'}{2T}]$, thereby enabling the possibility of sampling the corresponding signal at the receiver using the Nyquist rate of $\frac{M'}{T}$. 

Finally, 
by concatenating $N$ consecutive OFDM signals using a time division multiplexer, the time domain signal for OTFS in the coexisting system is obtained as  
\begin{align}\label{st}
x(t)&=\sum _{n=0}^{N-1}x_n\left(t-nT-\!\!\sum\limits_{\ddot{n}=0}^{n}a_{\ddot{n}}T^{\mathrm{cp}}_{\ddot{n}}\right),
\end{align}
where
$a_{\ddot{n}}=1$, if $ \ddot{n}\geqslant 1$ 
 and $a_{\ddot{n}}=0$, elsewhere.  

\begin{remark} \label{Rem:SelectionofMMprim}
We note that hardware implementation of $\Psi$-point fast Fourier transform (FFT) and $\Psi$-point IFFT operations are easier when $\Psi$ is set to a power of two.
Considering this, OFDM systems typically set the total number of sub-carriers in an OFDM signal ($M'$) to a power of two, so that the $M'$-point Fourier operation given in~\eqref{st2} at the transmitter, and a similar operation at the receiver, can be implemented in hardware with minimal complexity~\cite{2023_ODDM_TCOM}. 
However, when OTFS coexists with OFDM systems, in addition to the $M'$-point Fourier operation given in~\eqref{st2}, two other Fourier operations should be performed even within the transmitter. Specifically, since the ISFFT in~\eqref{Equ:Xtf} can be viewed as two Fourier operations, an $M$-point Fourier operation from delay domain to frequency domain and an $N$-point Fourier operation from Doppler domain to time domain would need to be performed at the preprocessing stage of the transmitter. 
On one hand, it is indeed possible to set the number of Doppler elements within the OTFS signal ($N$) to be the power of two to ensure the easier hardware implementation of the $N$-point Fourier operation in ISFFT. 
On the other hand, while it would be desirable to set the number of delay elements within an OTFS signal ($M$) to be a power of two for easier implementation of the $M$-point Fourier operation in ISFFT, this may not be possible. This is because when OTFS coexists with OFDM systems, the number of delay elements within an OTFS signal would be made equal to the total number of loaded sub-carriers in an OFDM signal. Since the number of loaded sub-carriers in an OFDM signal is usually chosen to satisfy the bandwidth constraint while minimizing OOBE, $M$ may not necessarily be a power of two. 
We note that this challenge related to the $M$-point Fourier operation is inevitable when OTFS coexists with OFDM systems featuring ECU.\footnote{To address this challenge, one can set the number of delay elements within the OTFS signal ($M$) to be lower than the number of loaded sub-carriers specified in the 3GPP standards, while ensuring $M$ is a power of two. However, this modification will come at the cost a reduction in data rate.}

\end{remark}

\subsubsection{OTFS with CPs of unequal length}

According to the frame structure that is defined in the 3GPP standards, the downlink and uplink transmissions are organized into \textit{radio frames}~\cite{3GPPStandOFDM,2018_ComStandardMag_OFDMNumerology,5GOFDMWebpage}. Then the radio frames are divided into \textit{sub-frames}. Finally, \textit{sub-frames} are further divided into two equally-sized half-frames of 0.5 ms, which we refer to as  \textit{time windows}. The time windows contain multiple OFDM signals~\cite[Chapter 4.1-4.3]{3GPPStandOFDM}.\footnote{It is noted that according to 3GPP standards, time windows are divided (or merged) to form \textit{slots} which carry OFDM signals~\cite[Chapter 4.1-4.3]{3GPPStandOFDM}. Scheduling is done on a subframe basis for both the downlink and uplink transmissions.}
Within each time window, all OFDM signals, except the first, are given the same CP length. 
The first OFDM signal within each time window is given a CP of longer length to ensure that an integer number of OFDM signals are within each time window~\cite[Chapter 5.3.1]{3GPPStandOFDM}.\footnote{According to 3GPP standards, the first and the $(7{\times}2^{\xi}+1)$th OFDM signals within each subframe are given a CP of longer length, where $\xi$ is the numerology~\cite[Chapter 5.3.1]{3GPPStandOFDM}. This implies that the first OFDM signal within each time window is given a CP of longer length.}
Table~\ref{tab:OFDMNum} shows $\Delta f$, the number of OFDM signals carried by time windows ($S$), CP length of regular OFDM signals ($T_{\mathrm{reg}}^{\mathrm{cp}}$), and CP length of the first OFDM signal in each time window ($T_{\mathrm{long}}^{\mathrm{cp}}$) adopted for different 5G New Radio (NR) numerologies (subcarrier spacing configurations)~\cite{2018_ComStandardMag_OFDMNumerology,3GPPStandOFDM,5GOFDMWebpage}.\footnote{\textcolor{black}{According to 3GPP standards, there are two types of CPs used for OFDM signals, namely ``normal CP'' and ``extended CP''~\cite[Section 4.2]{3GPPStandOFDM}. Except when the subcarrier spacing is 60 kHz, CPs of type "normal CP" are always utilized. We clarify that when CPs of type ``normal CP'' is used, OFDM signals are assigned with CPs of unequal/different lengths~\cite[Section 5.3.1]{3GPPStandOFDM}.}}

\begin{table}[t]
\footnotesize 
\caption{Illustration of $\Delta f$, S, $T_{\mathrm{reg}}^{\mathrm{cp}}$, and $T_{\mathrm{long}}^{\mathrm{cp}}$  adopted for 5G NR Numerologies~\cite{3GPPStandOFDM}.}
\begin{center}
\begin{tabular}{|p{0.1\linewidth}|p{0.125\linewidth}|p{0.17\linewidth}|p{0.15\linewidth}|p{0.15\linewidth}|p{0.15\linewidth}|}
\hline Numer- -ology $({\xi})$ & Subcarrier spacing ($\Delta f$) & OFDM symbols per time window~(S) & CP of regular symbols ($T_{\mathrm{reg}}^{\mathrm{cp}}$)& CP of long symbols ($T_{\mathrm{long}}^{\mathrm{cp}}$)\\
\hline 0 & 15 kHz & 7  & $4.69 \mu \mathrm{s}$ &   $5.2 \mu \mathrm{s}$  \\
\hline 1 & 30 kHz &  14   & $2.34 \mu \mathrm{s}$ & $2.86 \mu \mathrm{s}$  \\
\hline 2 & 60 kHz & 28     & $1.17 \mu \mathrm{s}$  & $1.69 \mu \mathrm{s}$\\
\hline 3 & 120 kHz & 56     & $0.59 \mu \mathrm{s}$ & $1.11 \mu \mathrm{s}$ \\
\hline 4 & 240 kHz & 112   & $0.29 \mu \mathrm{s}$  & $0.81 \mu \mathrm{s}$ \\
\hline
\end{tabular}\label{tab:OFDMNum}
\end{center}
\end{table}

It can be observed from Table~\ref{tab:OFDMNum} that the number of OFDM signals carried by a time window is typically small, especially for numerologies with low to moderate subcarrier spacing. On the other hand, 
to achieve reasonable time diversity, the number of Doppler elements within the OTFS signal, which is equal to the number of OFDM signals within the OTFS signal, may be set to be a reasonably high value. Due to these, the utilization of multiple time windows may become necessary to transmit an OTFS signal using an OFDM system.
\textit{This necessitates the investigation of OTFS modulation with CPs of unequal lengths.}
To account for the unequal lengths of CPs, in this work we consider $T^{\mathrm{cp}}_{n}$ to be 
\begin{align}\label{st3}
T^{\mathrm{cp}}_{n}&=\begin{cases}
T_{\mathrm{long}}^{\mathrm{cp}}, & ~~~~~~[n]_{S}=0,\\
T_{\mathrm{reg}}^{\mathrm{cp}}, & ~~~~~~\textrm{elsewhere},
\end{cases}
\end{align}
where $T_{\mathrm{long}}^{\mathrm{cp}}> T_{\mathrm{reg}}^{\mathrm{cp}}$ and $[.]_S$ denotes the mod $S$ operation.

\subsubsection{Terminology used in the Paper}
\label{Sec:Terminology}

The OTFS modulation in which a single CP is included in the entire OTFS signal is referred in the literature as RCP-OTFS.\footnote{\textcolor{black}{We note that the vector OFDM (VOFDM) was proposed in~\cite{2000_TCOM_VOFDM} for interference (ISI) channels that have spectral nulls. The VOFDM can be considered as a general transmission scheme, since OFDM and single-carrier frequency domain equalization (SC-FDE) are two special/extreme cases of VOFDM. It is interesting to note that VOFDM has the same digital sequence as that of RCP-OTFS
~\cite{2022TWC_CommentsonOTFSandVOFDMaretheSame}. Despite this, we clarify that the exploration of full diversity over doubly selective channels and coupling the information-bearing symbols over the DD domain representation of the channel were first discussed for RCP-OTFS, thereby highlighting the uniqueness of RCP-OTFS~\cite{2017_WCNC_OTFS_Haddani}. 
}} Also,  
the OTFS modulation in which multiple CPs are included in the OTFS signal is referred to as CP-OTFS or precoded-OFDM~\cite{2022_IEEEAccess_Viterbo_GeneralIOR,2022_EmanuelBook_DDCom,2018_PIMRC_IterativeDetectionforDSChannel}.
Based on these definitions, the variant of OTFS that is considered in this work falls into the category of CP-OTFS. However, it is worth noting that the variant of CP-OTFS considered in this work differs from those explored in the literature in two ways. The first characteristic pertains to whether the CPs added to the OTFS signal are of equal or unequal lengths. The second characteristic pertains to whether ECU is taken into consideration or not.  Thus, considering these differences, we will consistently use the following definitions throughout the paper to differentiate between the different variants of CP-OTFS:

\begin{itemize}
\item  \textit{CP-OTFS with CP of unequal length and ECU (CP-OTFS-w-UCP-ECU):} The variant of CP-OTFS in which (i) the CPs within the CP-OTFS signal are of different lengths and (ii) ECU is considered. This is the variant of CP-OTFS that is implementable on current 4G/5G communication systems;
\item  \textit{CP-OTFS with CPs of the unequal lengths (CP-OTFS-w-UCP):} The variant of CP-OTFS in which the CPs within the CP-OTFS signal are of different lengths, but with no ECU. This is the variant of OTFS that is implementable on OFDM systems that do not consider ECU;
\item  \textit{CP-OTFS with CPs of the same lengths (CP-OTFS-w-ECP):} The variant of CP-OTFS in which all the CPs within the CP-OTFS signal are of equal length. This is the variant of CP-OTFS that is explored to date in the literature\cite{2018_WCL_ModemStructureforOFDMBasedOTFS,2018_ICC_OTFSwithOFDM,2022_IEEEAccess_Viterbo_GeneralIOR,2019_ICC_CPOTFSwithMassiveMIMO}.
\end{itemize}

\noindent
It is interesting to that CP-OTFS-w-UCP and CP-OTFS-w-ECP can be considered as special cases of CP-OTFS-w-UCP-ECU. Moreover, CP-OTFS-w-ECP can also be considered as a special case of CP-OTFS-w-UCP.

\subsection{Channel}

When the signal is passed through a doubly-selective channel, the complex-valued baseband received signal becomes~\cite{2022_EmanuelBook_DDCom}
\begin{align} y(t) = \int_{\nu} {\int_{\tau} h(\nu,\tau)~e^{j2\pi \nu (t - \tau)}x(t - \tau) d\tau} d\nu + w(t), \label{Equ:RxSignal}
\end{align}
where $h(\nu,\tau)$ denotes the spreading function of the channel with $\nu$ and $\tau$ representing the Doppler and delay variables, respectively, and $w(t)\sim\mathcal{CN}(0,\sigma_n^2)$ denotes the
complex  noise. 
Considering that the  channel is composed of $I$  separable propagations paths, $h(\nu,\tau)$ can be represented as 
\begin{align}\label{Equ:ddchannel}
  h(\nu,\tau)=\sum_{i=1}^I h_i\delta(\nu-\nu_i)\delta(\tau-\tau_i),
\end{align}
where $h_i$, $\nu_i$, and $\tau_i$ denote the complex gain, Doppler, and delay of the $i$th propagation path, respectively, and $\delta(\cdot)$ denotes the Kronecker delta function. 
For simplicity, we consider that the Doppler and delay resolutions used to discretize the channel in the DD domain are sufficiently small  such that Dopplers and delays of propagation paths can be approximated to their nearest on-grid values, i.e., on-grid Dopplers and on-grid delays (a.k.a integer Dopplers and integer delays~\cite{2018_TWC_Viterbo_OTFS_InterferenceCancellation,2022_EmanuelBook_DDCom,2020_TVT_Viterbo_RakeDFEforOTFS}). Considering this, we express $\nu_i$ and $\tau_i$ as 
\begin{align} \label{Equ:lpkp}
  \nu_i =\frac{k_i}{NT},~~~~~ \tau_i =\frac{T l_i}{M'}, ~~~~~~~\forall~ i\in\mathcal{I}\in{\{1,2,..., I\}},
\end{align}
where $k_i$ and $l_i$ are integers representing the Doppler and delay indices 
for the $i$th propagation path, respectively. 

\textcolor{black}{We note that the duration of the considered CP-OTFS signal given in~\eqref{st} is $NT+\!\!\sum\limits_{\ddot{n}=0}^{N}a_{\ddot{n}}T^{\mathrm{cp}}_{\ddot{n}}$.
This would imply that the resolution used to discretize the doubly-selective channel in the Doppler domain could be $1/(NT+\!\!\sum\limits_{\ddot{n}=0}^{N}a_{\ddot{n}}T^{\mathrm{cp}}_{\ddot{n}})$. However, 
we clarify that 
despite RCP-OTFS signal duration being $NT+T^{\mathrm{cp}}$, all previous studies on RCP-OTFS  have employed $\frac{1}{NT}$ as the Doppler resolution~\cite{2022_EmanuelBook_DDCom,2020_TVT_Viterbo_RakeDFEforOTFS,2018_TWC_Viterbo_OTFS_InterferenceCancellation}. Considering this and to be consistent with previous studies on CP-OTFS in~\cite{2022_IEEEAccess_Viterbo_GeneralIOR,2022_EmanuelBook_DDCom,akramGC2023_OTFSwithOFDM}, we adopt the resolution $\frac{1}{NT}$ in~\eqref{Equ:lpkp}  for discretizing the channel in the Doppler domain.}\footnote{\textcolor{black}{Despite our consideration, we note that the derivations in this work can be directly extended  even if  an alternative value, such as $1/(NT+\!\!\sum\limits_{\ddot{n}=0}^{N}a_{\ddot{n}}T^{\mathrm{cp}}_{\ddot{n}})$, is adopted for the Doppler resolution.}}

Moreover, since OFDM receivers employ a sampling rate of $\frac{M'}{T}$ to discretize $y(t)$, in~\eqref{Equ:lpkp}, we adopt  the resolution $\frac{T}{M'}$  for  discretizing the channel in the delay domain. It is interesting to note that due to the presence of ECU, 
the sampling rate used to discretize $y(t)$ is different from the effective bandwidth  occupied by $y(t)$, which is $\frac{M}{T}$.

Finally, substituting \eqref{Equ:ddchannel} and~\eqref{Equ:lpkp} in \eqref{Equ:RxSignal}, we obtain 
\begin{align} \label{Equ:rt}
y(t) &= \sum_{i=1}^I h_ie^{j\frac{2\pi k_i}{NT}(t - \frac{T l_i}{M'})}x(t - \frac{T l_i}{M'})+ w(t).
\end{align}

\subsection{Receiver}

In practical OFDM receivers, the received signal $y(t)$ is first passed through a band-pass filter 
to obtain the filtered signal $\tilde{y}(t)$. Then, $\tilde{y}(t)$ is sampled in the time domain at the sampling rate of $\frac{M'}{T}$. 
Then, the samples corresponding to the CPs are appropriately removed from the sampled time domain signal.
Mathematically, the $\ell$th sample of the sampled time domain signal after CP removal is expressed as

\begin{align} \label{Equ:el}
\tilde{y}[\ell]=\int_{-\infty}^{\infty}\tilde{y}(t)\delta \Big(t-\big(\frac{T\ell}{M'}{+}\!\!\sum\limits_{\ddot{n}=0}^{\lfloor \ell/M'\rfloor}\!\!a_{\ddot{n}}T^{\mathrm{cp}}_{\ddot{n}}\big)\Big)dt,
\end{align}
 where $\lfloor \cdot\rfloor$ denotes the floor operation. 

Next, $\tilde{y}[\ell]$ is sent through a series-to-parallel (S/P) converter and a FFT operation to obtain the received extended TF domain signal. Specifically, the $(n,m)$-th symbol of the received extended TF domain signal,  $Y'_{\textrm{TF}}[n,m]$, is obtained as
\begin{align}
&\!\!\!\!\!\!Y'_{\textrm{TF}}[n,m] 
{=}\frac{1}{\sqrt{M'}}\!\!\!\sum_{s=0}^{M'-1}\tilde{y}[nM'{+}s]e^{{-}j\frac{2 \pi m s}{M'}},
\label{Equ:Ymn2}
\end{align}
where $n\in\mathcal{N},m\in\mathcal{M}'$. 
Then, disregarding the symbols that correspond to  the unloaded edge carriers,  the received TF domain signal is obtained from $Y'_{\textrm{TF}}$. Specifically, the $(n,m)$-th symbol of the received TF domain signal,  $Y_{\textrm{TF}}[n,m]$, is obtained as 
\begin{align}\label{Equ:Ymn}
&Y_{\textrm{TF}}[n,m]=\begin{cases}
Y_{\textrm{TF}}'[n,m], & 0 \leqslant m < \frac{M}{2}\\
Y_{\textrm{TF}}'[n,m-M], & \frac{M}{2} \leqslant m \leqslant M-1 ,\\
\end{cases}
\end{align}
where $n\in\mathcal{N},m\in\mathcal{M}$.

At the receiver postprocessing stage, the received DD domain signal is obtained by applying simplectic finite Fourier transform (SFFT) on $Y_{\textrm{TF}}$. Specifically, the $(k,l)$-th symbol of  the received DD domain signal, $Y_{\textrm{DD}}[k,l]$, is obtained as 
\begin{align}\label{Equ:ylk1}
    &Y_{\textrm{DD}}[k,l] = \frac{1}{\sqrt{NM}} \sum _{n=0}^{N-1} \sum _{m=0}^{M-1} Y_{\textrm{TF}}[n,m] e^{-j2\pi \left({\frac{nk}{N}-\frac{ml}{M}}\right)},
\end{align}
where $k\in\mathcal{N},l\in\mathcal{M}$. Finally, the signal detection and channel estimation are performed based on $Y_{\textrm{DD}}$.

\section{Input-Output Relation}\label{section:3}

In this section, we characterize the IOR for CP-OTFS-w-UCP-ECU to understand the coexistence of OTFS modulation with OFDM systems. 

\begin{figure*}[!t]
\normalsize 
\begin{align}
\tilde{y}[\ell] & 
= \frac{1}{\sqrt{M'}}\sum_{i=1}^I \sum _{\bar{n}=0}^{N-1}\sum _{\bar{m}=-\frac{M'}{2}}^{\frac{M'}{2}-1}h_i X'_{\textrm{TF}}[\bar{n},\bar{m}]e^{j\frac{2\pi}{M'}(\ell-l_i)\left(\bar{m} +\frac{k_i}{N}\right) } e^{\!\!-j2\pi\bar{m} \left( \bar{n}{+}\!\!\sum\limits_{\ddot{n}=0}^{\bar{n}}\!a_{\ddot{n}}\psi_{\ddot{n}}^{\mathrm{cp}}{-}\!\!\!\!\sum\limits_{\ddot{n}=0}^{\lfloor \ell/M'\rfloor}\!\!a_{\ddot{n}}\psi_{\ddot{n}}^{\mathrm{cp}}\right)} e^{j\frac{2\pi k_i}{N}\!\!\sum\limits_{\ddot{n}=0}^{\lfloor \ell/M'\rfloor}\!a_{\ddot{n}}\psi_{\ddot{n}}^{\mathrm{cp}}}\notag\\[-6pt]
&~~~~~~~~~~~~~~~~~~~~~~~~~~~~~~~~~~~~~~~~~~~~~~~~~~~~~~~~~~~~~~~\times g_{\bar{n}}(\frac{T}{M'}(\ell{-}l_i{-}M'(\bar{n}{+}\!\!\sum\limits_{\ddot{n}=0}^{\bar{n}}\!a_{\ddot{n}}\psi_{\ddot{n}}^{\mathrm{cp}}{-}\!\!\!\!\sum\limits_{\ddot{n}=0}^{\lfloor \ell/M'\rfloor}\!\!a_{\ddot{n}}\psi_{\ddot{n}}^{\mathrm{cp}}))), \label{Equ:ri2}\\[-4pt] 
Y'_{\textrm{TF}}[n,m] 
&=\frac{1}{M'}\sum_{s=0}^{M'-1}\sum_{i=1}^I \sum _{\bar{m}=-\frac{M'}{2}}^{\frac{M'}{2}-1}h_iX'_{\textrm{TF}}[n,\bar{m}]e^{j\frac{2\pi k_i}{N}(n+\!\!\sum\limits_{\ddot{n}=0}^{n}\!a_{\ddot{n}}\psi_{\ddot{n}}^{\mathrm{cp}})}e^{-j\frac{2\pi}{M'} \left( (m-\bar{m}-\frac{k_i}{N})s+l_i(\bar{m}+\frac{k_i}{N})\right)}.
\label{Equ:Ymn3}\\[-2pt]
Y_{\textrm{TF}}[n,m]
&=\frac{1}{M'}\sum_{s=0}^{M'{-}1}\sum_{i=1}^I \sum_{\bar{m}=0}^{M{-}1}h_iX_{\textrm{TF}}[n,\bar{m}]e^{j\frac{2\pi k_i}{N}(n+\!\!\sum\limits_{\ddot{n}=0}^{n}\!a_{\ddot{n}}\psi_{\ddot{n}}^{\mathrm{cp}})}e^{{-}j\frac{2\pi}{M'} \left( (m{-}\bar{m}{-}\frac{k_i}{N})s+l_i(\bar{m}+\frac{k_i}{N})\right)} \Phi_{m,\bar{m}}(s). \label{Equ:Ymn4} 
\end{align}
\hrulefill
\end{figure*}

When the received signal $y(t)$ is  bandlimited, the filtered signal of  $y(t)$ can be written as $\tilde{y}(t)=y(t)$.\footnote{If we consider that the channel model given in~\eqref{Equ:ddchannel} represents the sampled equivalent channel after undergoing both bandpass filtering and sampling, with $\nu_i$ and $\tau_i$ corresponding to~\eqref{Equ:lpkp}, we can regard $y(t)$ in~\eqref{Equ:rt} as the received signal that results following the bandpass filtering process.} Based on this, while utilizing~\eqref{Equ:rt}, we simplify $\tilde{y}[\ell]$ in~\eqref{Equ:el} in the absence of noise
to obtain~\eqref{Equ:ri2}.
The expression~\eqref{Equ:ri2}, along with~\eqref{Equ:Ymn3} and~\eqref{Equ:Ymn4}, is given at the start of the next page. In~\eqref{Equ:ri2}, $\psi_{n}^{\mathrm{cp}}=\frac{T_{n}^{\mathrm{cp}}}{T}$.
Then substituting~\eqref{Equ:ri2} in~\eqref{Equ:Ymn2} while leveraging the fact that $g_{\bar{n}}(\frac{T}{M'}(s-l_i-M'(\bar{n}-n{+}\!\!\sum\limits_{\ddot{n}=0}^{\bar{n}}\!a_{\ddot{n}}\psi_{\ddot{n}}^{\mathrm{cp}}-\!\!\sum\limits_{\ddot{n}=0}^{n}\!\!a_{\ddot{n}}\psi_{\ddot{n}}^{\mathrm{cp}})))$ is non-zero $\forall s\in\mathcal{M}'$ only when $\bar{n}=n$, we obtain $Y'_{\textrm{TF}}[n,m]$  in terms of $X'_{\textrm{TF}}[n,m]$ as~\eqref{Equ:Ymn3}. 
\noindent
We next substitute~\eqref{Equ:Ymn3} and~\eqref{Equ:Xtfprim} in~\eqref{Equ:Ymn} and rearrange the terms to obtain $Y_{\textrm{TF}}[n,m]$ in terms of $X_{\textrm{TF}}[n,m]$ as~\eqref{Equ:Ymn4}, where $\Phi_{m,\bar{m}}(s)$ in it is given by
\begin{align}\label{Equ:phimmbars}
  & \Phi_{m,\bar{m}}(s)    \notag\\
  &= \begin{cases}
1,& 0~\!{\leqslant}~\! m,~\bar{m} ~\!{<}~\! \frac{M}{2}\\
e^{-\frac{j2\pi}{M'}(M'{-}M)({-}s{+}l_i)} ,& 0\leqslant m < \frac{M}{2},~\bar{m} \geqslant \frac{M}{2},\\
e^{-\frac{j2\pi}{M'}(M'{-}M)s},& m \geqslant \frac{M}{2},~0~\!{\leqslant}~\! \bar{m} < \frac{M}{2},\\
e^{-\frac{j2\pi}{M'}(M'{-}M)l_i},& m,~\bar{m} \geqslant \frac{M}{2},
\end{cases}
\end{align}
where $m,\bar{m}\in\mathcal{M}$ and $s\in\mathcal{M'}$.
\noindent
We next substitute~\eqref{Equ:Ymn4} and~\eqref{Equ:Xtf} in~\eqref{Equ:ylk1} and rearrange the terms to 
arrive at 
\begin{align}\label{Equ:ylk2}
    Y_{\textrm{DD}}[k,l]     &= \sum_{\bar{k}=0}^{N-1} \sum_{\bar{l}=0}^{M-1}  h_{w}(k,\bar{k},l,\bar{l}) X_{\textrm{DD}}[\bar{k},\bar{l}],
\end{align}
where 
\begin{align}\label{Equ:hw}
    h_{w}(k,\bar{k},l,\bar{l})  
&=\sum_{i=1}^I  h_i e^{{-}j\frac{2 \pi l_ik_i}{M'N}}\mathcal{\tilde{G}}(k,\bar{k},k_i)\mathcal{\tilde{F}}(l,\bar{l},l_i,k_i),\\
\label{Equ:G}
   \mathcal{\tilde{G}}(k,\bar{k},k_i)&=\frac{1}{N}\sum _{n=0}^{N-1} e^{-j\frac{2 \pi }{N}(n(k-\bar{k}-k_i)-k_i\!\!\sum\limits_{\ddot{n}=0}^{n}a_{\ddot{n}}\psi_{\ddot{n}}^{\mathrm{cp}})},\\
   \!\!\mathcal{\tilde{F}}(l,\bar{l},l_i,k_i)&\!=\frac{1}{MM'}\!\!\!\sum_{s=0}^{M'-1}\sum_{\bar{m}=0}^{M-1}\sum _{m=0}^{M-1}\!\! e^{{-}j\frac{2\pi}{M'} \left( (m{-}\bar{m}{-}\frac{k_i}{N})s+l_i\bar{m}\right)} \notag\\
&~~~~~~~~~~~~~~~~~~~\times e^{-j\frac{2\pi}{M}(\bar{m}\bar{l}+ml)}\Phi_{m,\bar{m}}(s). \!\!\! \label{Equ:F}
\end{align}

\noindent
\textcolor{black}{Thereafter, we further simplify~\eqref{Equ:G} and~\eqref{Equ:F} and then substitute them in~\eqref{Equ:hw}. Finally,  we substitute~\eqref{Equ:hw} in~\eqref{Equ:ylk2} to arrive at the IOR for CP-OTFS-w-UCP-ECU. The result is presented in the following theorem.}

\begin{figure*}[!t]
\normalsize 
\begin{align}\label{Equ:ylk3}
    Y_{\textrm{DD}}[k,l]&=\sum_{i=1}^I h_i e^{{-}j\frac{2 \pi l_ik_i}{M'N}} \!\!\sum_{q={-}\frac{N}{2}}^{\frac{N}{2}{-}1}\!\!\mathcal{G}(q, k_i)  \!\!\sum_{\ell={-}\frac{M}{2}}^{\frac{M}{2}{-}1}  \!\!\mathcal{F}(\ell,l,l_i,k_i) X_{\textrm{DD}}\left[[k{-}k_i{+}q]_N,[l{-}|| l_i/\mu|| {+}\ell]_M\right],\\
   \mathcal{G}(q,k_i)&=\begin{cases}
\delta(q),& k_i=0,\\
\frac{e^{-j\frac{2 \pi S \omega_{\textrm{f}}}{N} (-q-k_i(\psi^{\mathrm{reg}}+\frac{\psi^{\mathrm{ext}}}{S}))}-1}{e^{-j\frac{2 \pi S}{N}(-q-k_i(\psi^{\mathrm{reg}}+\frac{\psi^{\mathrm{ext}}}{S}))}-1} \frac{e^{-j\frac{2 \pi S}{N}(-q-k_i\psi^{\mathrm{reg}})}-1}{N e^{-j\frac{2 \pi }{N}(-q-k_i\psi^{\mathrm{reg}})}-N}\\
~~~~~~~~~~~-e^{-j\frac{2\pi S(\omega_{\textrm{f}}-1)}{N} (-q-k_i(\psi^{\mathrm{reg}}+\frac{\psi^{\mathrm{ext}}}{S}))}\frac{e^{-j\frac{2 \pi S}{N}(-q-k_i\psi^{\mathrm{reg}})}-e^{-j\frac{2 \pi \omega_{\textrm{m}}}{N}(-q-k_i\psi^{\mathrm{reg}})}}{N e^{-j\frac{2 \pi }{N}(-q-k_i\psi^{\mathrm{reg}})}-N},& k_i\neq0,
\end{cases} \label{Equ:G5}\\
     \mathcal{F}(\ell,l,l_i,k_i) &=\begin{cases}
\delta(\ell),& k_i=[l_i]_\mu=0,\\
\frac{e^{-j \pi (-\ell-\lfloor \frac{l_i}{\mu}\rfloor)}-1}{Me^{-j\frac{2 \pi }{M}(-\ell-\lfloor \frac{l_i}{\mu}\rfloor)}-M}(1+e^{-j\frac{2 \pi (M'-M)l_i }{M'}}e^{-j \pi (-\ell-\lfloor \frac{l_i}{\mu}\rfloor)}),& k_i=0, [l_i]_\mu \neq 0,\\
\sum_{s=0}^{M'-1}e^{j\frac{2\pi sk_i}{NM'}}\frac{e^{-j \pi (\frac{s}{\mu}-l)}-1}{M'e^{-j\frac{2 \pi }{M}(\frac{s}{\mu}-l)}-M'}\frac{e^{-j \pi (\ell-\frac{s}{\mu}+l+\lfloor \frac{l_i}{\mu}\rfloor)}-1}{Me^{-j\frac{2 \pi }{M}(\ell-\frac{s}{\mu}+l+\lfloor \frac{l_i}{\mu}\rfloor)}-M},& \\
~~~~~~~~~\times (1+e^{-j\frac{2 \pi (M'-M)s }{M'}}e^{-j \pi (\frac{s}{\mu}-l)})(1+e^{j\frac{2 \pi (M'-M)(s-l_i) }{M'}}e^{-j \pi (\ell-\frac{s}{\mu}+l+\lfloor \frac{l_i}{\mu}\rfloor)}),& [l_i]_\mu \neq 0.
\end{cases} \label{Equ:F_Fin}\!\!\!
\end{align}
\hrulefill
\end{figure*}

\begin{theorem}\label{Thr:IORCPOTFS-VCP-ECU}
The DD domain received signal $Y_{\textrm{DD}}[k,l]$  for CP-OTFS-w-UCP-ECU in the absence of noise is given by~\eqref{Equ:ylk3}, shown at the start of this page. In~\eqref{Equ:ylk3}, 
$\mathcal{G}(q,k_i)$ and $\mathcal{F}(\ell,l,l_i,k_i)$ denote the spreading function along the Doppler and delay dimensions, respectively, and are obtained as~\eqref{Equ:G5} and~\eqref{Equ:F_Fin}, which are shown at the start of this page. In~\eqref{Equ:ylk3}-\eqref{Equ:F_Fin}, $\psi^{\mathrm{reg}}{=}\frac{T_{\mathrm{reg}}^{\mathrm{cp}}}{T}$, $\psi^{\mathrm{ext}}{=}\frac{T_{\mathrm{long}}^{\mathrm{cp}}-T_{\mathrm{reg}}^{\mathrm{cp}}}{T}$, $\omega_{\textrm{f}}=\lceil\frac{N}{S}\rceil$ denotes the number of time windows required to transmit an OTFS signal, $\omega_{\textrm{m}}=[N]_S$, $\mu=\frac{M'}{M}$, and $\lceil\cdot\rceil$ and $||\cdot||$ denote the ceil and round operations, respectively.

\textit{Proof:} The steps followed to obtain~\eqref{Equ:ylk3} from~\eqref{Equ:ylk2} are relegated to Appendix~\ref{App:IORCPOTFS-VCP-ECU}.  \hfill $\blacksquare$
\end{theorem}

To gain a clearer understanding of the impacts of the inclusion of multiple CPs to the OTFS signal and ECU,  prior to interpreting the results presented in Theorem 1, we simplify Theorem 1 focusing on the two special cases of the considered system model, which were discussed in Section~\ref{Sec:Terminology}. 
These simplifications will distinctly illustrate the impacts resulting from our considerations.

\subsection{Special Cases}

\subsubsection{Special Case~1:CP-OTFS-w-UCP}
By considering $M'{=}M$ in Theorem~\ref{Thr:IORCPOTFS-VCP-ECU}, we obtain the IOR of CP-OTFS-w-UCP and present it in the following Corollary.

\begin{corollary}\label{Cor:IORVCPOTFS}
The DD domain  received signal $Y_{\textrm{DD}}[k,l]$ for CP-OTFS-w-UCP in the absence of noise is given by 
\begin{align}\label{Equ:ylkSC}
    Y_{\textrm{DD}}[k,l] &= \sum_{i=1}^I h_i e^{j\frac{2 \pi (l-l_i)k_i}{NM}} \sum_{q=-\frac{N}{2}}^{\frac{N}{2}-1} \mathcal{G}(q, k_i)  \notag\\
&~~~~~~~~~~~~~\times X_{\textrm{DD}}\left[[k-k_i+q]_N,[l-l_i]_M\right].
\end{align}
\noindent
\end{corollary}

\subsubsection{Special Case~2: CP-OTFS-w-ECP}

We simplify  Corollary~\ref{Cor:IORVCPOTFS} by considering $T_{\mathrm{long}}^{\mathrm{cp}}=T_{\mathrm{reg}}^{\mathrm{cp}}$, or equivalently $\psi^{\mathrm{ext}}{=}0$, to obtain the IOR of CP-OTFS-w-ECP, and present it in the following Corollary. 

\begin{corollary}\label{Cor:IORCPOTFS}
The DD domain received signal $Y_{\textrm{DD}}[k,l]$ for CP-OTFS-w-ECP in the absence of noise is given by 
\begin{align}\label{Equ:ylkSC2}
    Y_{\textrm{DD}}[k,l] &= \sum_{i=1}^I h_i e^{j\frac{2 \pi (l-l_i)k_i}{NM}} \sum_{q=-\frac{N}{2}}^{\frac{N}{2}-1} \mathcal{G}_{\textrm{SC}}(q,k_i)  \notag\\
&~~~~~~~~~~~~~\times X_{\textrm{DD}}[[k-k_i+q]_N,[l-l_i]_M],
\end{align}
where 
\begin{align}\label{Equ:Gsc}
    \mathcal{G}_{\textrm{SC}}(q,k_i)=\begin{cases}
\delta(q),& ~~k_i=0,\\
\frac{e^{-j2\pi(-q-k_i \psi^{\mathrm{reg}})}-1}{N e^{-j\frac{2 \pi }{N}(-q-k_i \psi^{\mathrm{reg}})}-N},
& ~~\textrm{elsewhere}.
\end{cases}
\end{align}
\noindent
\end{corollary}

\subsection{Discussion}

In this subsection, we discuss the impacts of the inclusion of multiple CPs to the OTFS signal, unequal lengths of those CPs, and the consideration of ECU, using the results presented in Theorem~\ref{Thr:IORCPOTFS-VCP-ECU} and Corollaries~\ref{Cor:IORVCPOTFS} and~\ref{Cor:IORCPOTFS}.

From~\eqref{Equ:ylkSC} and~\eqref{Equ:ylkSC2} in Corollaries~\ref{Cor:IORVCPOTFS} and~\ref{Cor:IORCPOTFS}, respectively, as well as~\eqref{Equ:ylk3} in Theorem~\ref{Thr:IORCPOTFS-VCP-ECU},  we observe that even when Dopplers of all the propagation paths are on-grid, the inclusion of multiple CPs to the OTFS signal leads to the channel response exhibiting spreading effects/leakage along the Doppler dimension. In other words, every single propagation path is perceived as multiple taps that have the same delay and different Dopplers. 
This spreading effect along the Doppler dimension is characterized by $\mathcal{G}(q,k_i)$ for CP-OTFS-w-UCP-ECU and CP-OTFS-w-UCP, and by $\mathcal{G}_{\textrm{SC}}(q,k_i)$ for CP-OTFS-w-ECP.

By comparing~\eqref{Equ:ylk3} in Theorem~\ref{Thr:IORCPOTFS-VCP-ECU} with~\eqref{Equ:ylkSC} in Corollary~\ref{Cor:IORVCPOTFS}, we observe the presence of the spreading function along the delay dimensions, $\mathcal{F}(\ell,l,l_i,k_i)$, in Theorem~\ref{Thr:IORCPOTFS-VCP-ECU} but not in Corollary~\ref{Cor:IORVCPOTFS}. This shows that even when Dopplers and delays of all the propagation paths are on-grid, ECU  leads to the channel response exhibiting spreading effects/leakage along the delay dimension. Based on the above insights,
it can be concluded that \textit{the effective sampled DD domain channel model for OTFS in coexisting systems may exhibit reduced sparsity}.

By comparing~\eqref{Equ:ylkSC} in Corollary~\ref{Cor:IORVCPOTFS} with~\eqref{Equ:ylkSC2} in Corollary~\ref{Cor:IORCPOTFS}, we observe that the spreading function along the Doppler dimensions for CP-OTFS-w-UCP is different from that of CP-OTFS-w-ECP, i.e., $\mathcal{G}(q,k_i)\neq \mathcal{G}_{\textrm{SC}}(q,k_i)$. This shows that effective DD domain channel coefficients for OTFS in coexisting systems are influenced by the unequal lengths of the CPs. 

\textcolor{black}{
\begin{remark} \label{Rem:DopplerSpread}
It has been demonstrated that even when implementing OTFS modulation as a stand-alone system (such as the implementation of the RCP-OTFS variant), channel response can exhibit spreading effects/leakage along the Doppler and delay dimensions if the Dopplers and delays of the propagation paths are off-grid~\cite{2018_TWC_Viterbo_OTFS_InterferenceCancellation,2022_TWC_Zhigiang_OffGridCEforOTFS}. As a result, for OTFS in coexisting systems with off-grid Dopplers and delays, firstly, the combination of off-grid Doppler and the inclusion of multiple CPs to the OTFS signal causes the channel response to exhibit spreading effects along the Doppler dimension. Secondly, the combination of off-grid delay and ECU causes the channel response to exhibit spreading effects along the delay dimension.
\end{remark}}

\begin{figure*}[!t]
\centering\subfloat[\label{2a}RCP-OTFS]{ \includegraphics[width=0.8\columnwidth]{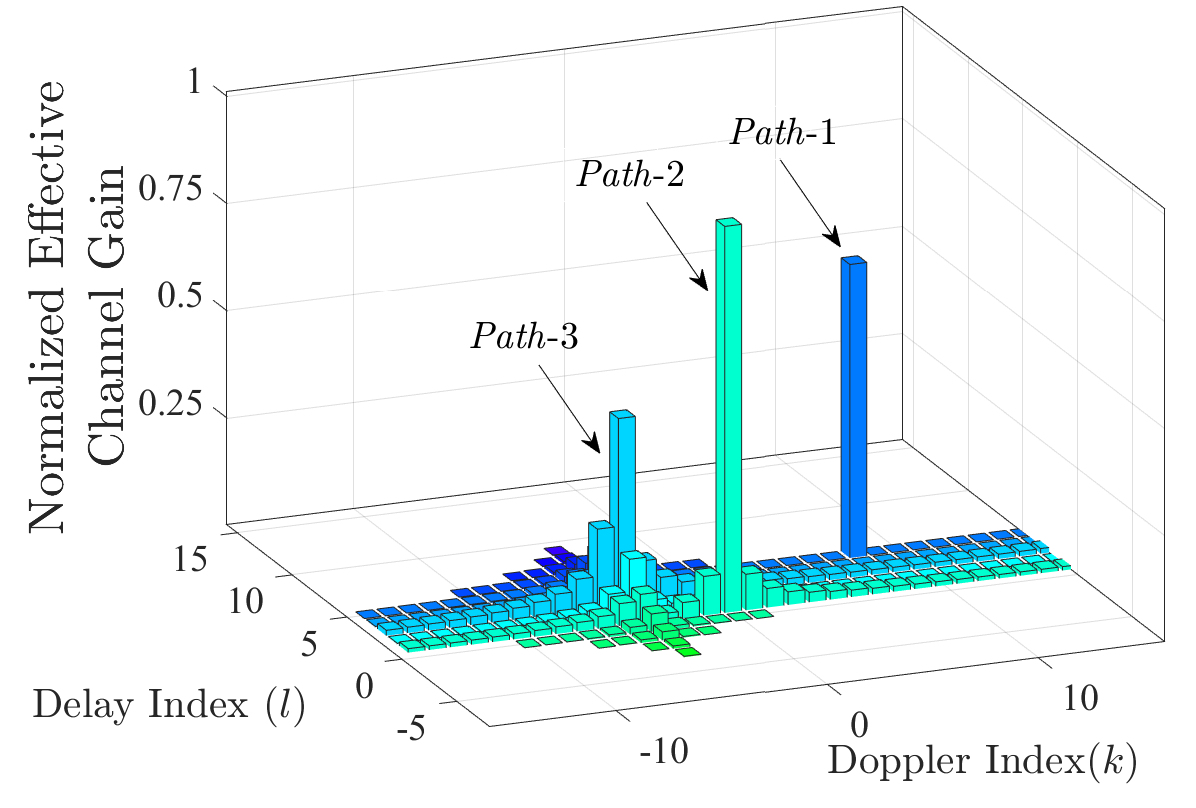}} 
\subfloat[\label{2b}CP-OTFS-w-ECP]{\includegraphics[width=0.8\columnwidth]{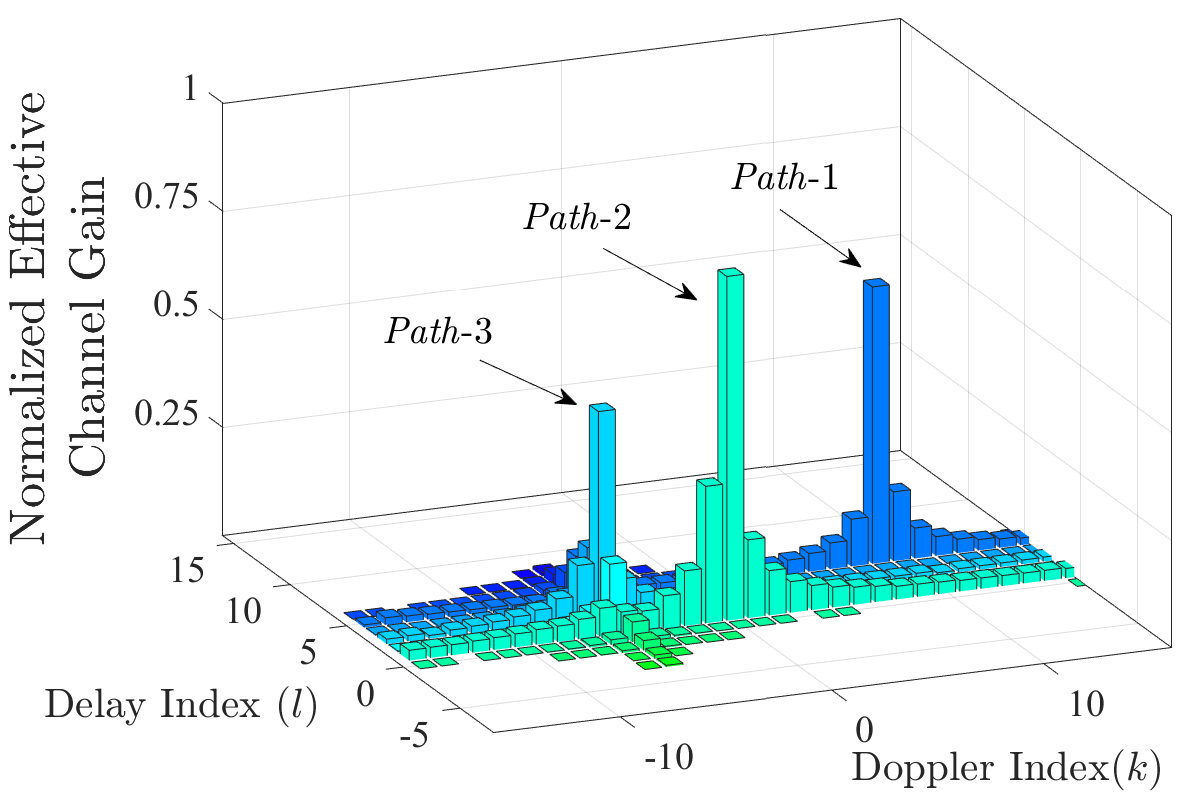}}
  \vspace{4 mm}
  \subfloat[\label{2a}CP-OTFS-w-UCP]{ \includegraphics[width=0.8\columnwidth]{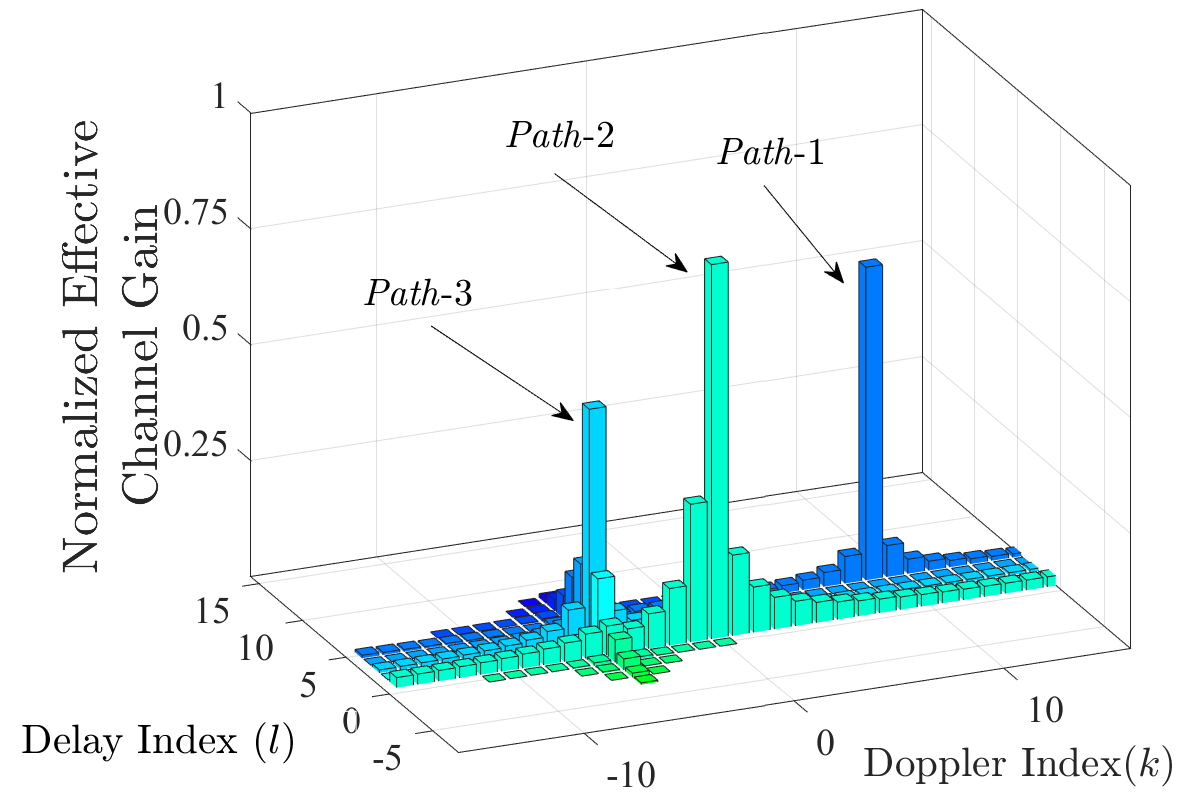}} \hspace{4 mm} 
  \subfloat[\label{2b}CP-OTFS-w-UCP-ECU]{\includegraphics[width=0.8\columnwidth]{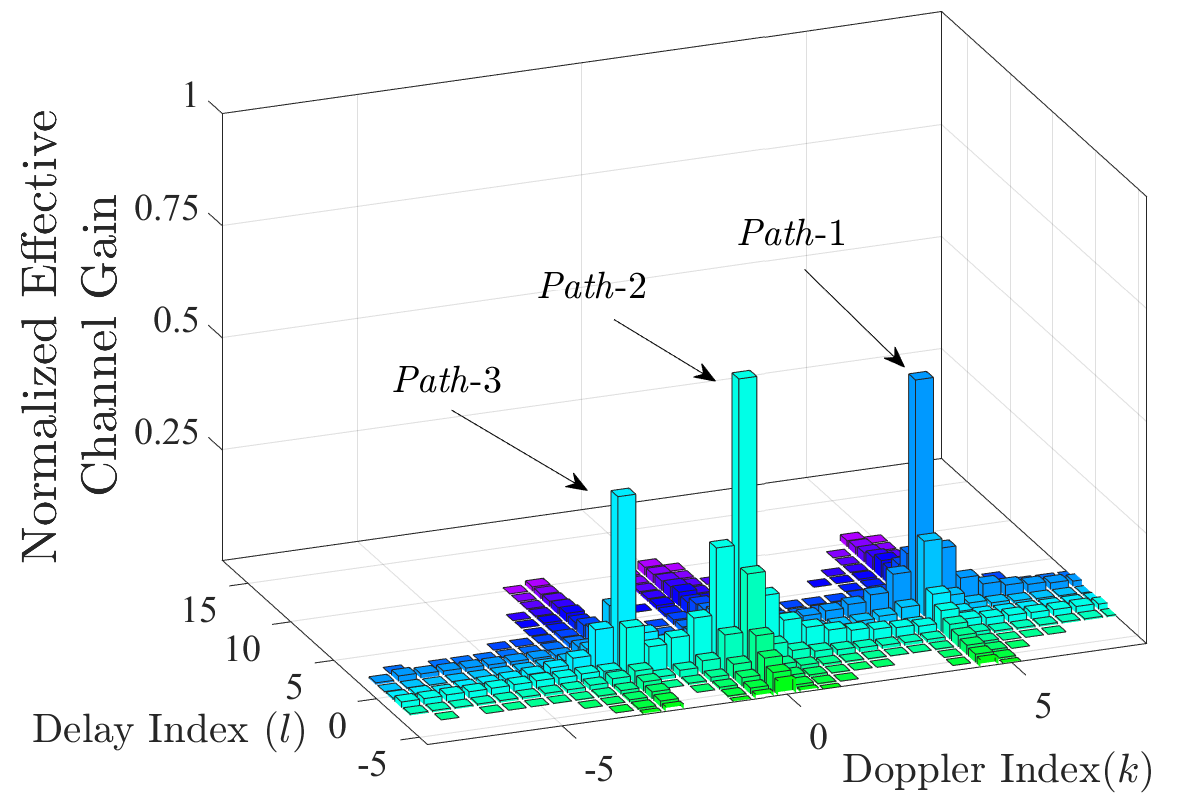}}
  \vspace{+3mm}
  \caption{Illustration of the normalized  effective sampled DD domain channel model for different variants of OTFS (Phase terms are ignored for the purpose of display).}\label{Fig:ChanGainCompare}
\end{figure*}

\subsubsection{Visual Illustration}

To visualize the  impacts of  the inclusion of multiple CPs to the OTFS signal and ECU, 
we plot the normalized  effective sampled DD domain channel model for RCP-OTFS, CP-OTFS-w-ECP, CP-OTFS-w-UCP, and CP-OTFS-w-UCP-ECU in Fig.~\ref{Fig:ChanGainCompare}. For this demonstration, we consider a 
doubly-selective channel with three propagation paths. 
The first propagation path is set to have on-grid Doppler and on-grid delay. The second path is set to have off-grid Doppler and on-grid delay, and the third path is set to have off-grid Doppler and off-grid delay. 
For convenience, consider the three paths to be denoted by \textit{Path-1}, \textit{Path-2}, and \textit{Path-3}, respectively. In Fig.~\ref{Fig:ChanGainCompare}, we set $N=M'=32$, and $M=24$.

We first observe from Fig.~\ref{Fig:ChanGainCompare}(a) that for RCP-OTFS, around the path with on-grid Doppler and on-grid delay (\textit{Path-1}), channel response does not exhibit spreading effects. Around the path that has off-grid Doppler and on-grid delay (\textit{Path-2}), channel response exhibits spreading effects only along the Doppler dimension. Moreover,  around the path that has off-grid Doppler and off-grid delay (\textit{Path-3}), channel response exhibits spreading effects along both Doppler and delay dimensions. These observations indicate that the channel model for RCP-OTFS can exhibit spreading effects along Doppler and/or delay dimension(s). However, this spreading occurs only when the propagation paths in the channel encounter off-grid Dopplers and/or off-grid delays.

Second, we observe from Figs.~\ref{Fig:ChanGainCompare}(b) and (c) that channel response exhibits spreading effects along the Doppler dimension for CP-OTFS-w-ECP and CP-OTFS-w-UCP around all their paths, regardless of whether the paths have on-grid Doppler or not.  The channel response exhibiting spreading effects along the Doppler dimension around the paths having on-grid Doppler, i.e., \textit{Path-1}, is a consequence of including multiple CPs in the OTFS signal. As for channel response exhibiting spreading effects around the paths having off-grid Doppler, i.e., \textit{Path-2} and \textit{Path-3}, it is a consequence of off-grid Doppler and the inclusion of multiple CPs to the OTFS signal.

Third, we observe from Fig.~\ref{Fig:ChanGainCompare}(d) that the channel response for CP-OTFS-w-ECP-ECU exhibits spreading effects along both the Doppler and delay dimensions around all their paths, regardless of whether the paths have on-grid delay or not.
The channel response exhibiting spreading effects along the delay dimension around the paths having on-grid delay, i.e., \textit{Path-1} and \textit{Path-2}, is a direct result of ECU. As for the channel response exhibiting spreading effects around the paths having off-grid delay, i.e., \textit{Path-3}, it is a consequence of off-grid delay and ECU.

Finally, by comparing  Figs.~\ref{Fig:ChanGainCompare}(b) with (c), we observe that the coefficients of the effective sampled channel response of CP-OTFS-w-UCP differ from CP-OTFS-w-ECP, due to the existence of CPs of unequal lengths in CP-OTFS-w-UCP.

\begin{remark} \label{Rem:GFApprox}
By analyzing~\eqref{Equ:G5} and~\eqref{Equ:F_Fin}, we find that the magnitudes of $\mathcal{G}(q,k_i)$ and $\mathcal{F}(\ell,l,l_i,k_i)$ have their maximum around $q=0$ and $\ell=0$, respectively. Also, they decrease significantly as $|q|$ and  $|\ell|$ increases, respectively, where $|.|$ denotes the absolute operation~\cite{2018_TWC_Viterbo_OTFS_InterferenceCancellation}. 
This behaviour of $\mathcal{G}(q,k_i)$ and $\mathcal{F}(\ell,l,l_i,k_i)$ is observed in Fig.~\ref{Fig:ChanGainCompare}(d) as well. 
Thus, by considering only a small number of terms in the summation on $q$ and $\ell$ in~\eqref{Equ:ylk3}, 
the IOR for CP-OTFS-w-UCP-ECU in Theorem 1 can be approximated as 
\begin{align}\label{Equ:ylk3_Approx}
    &Y_{\textrm{DD}}[k,l]\notag\\
    &=\sum_{i=1}^I h_i e^{{-}j\frac{2 \pi l_ik_i}{M'N}}  \!\!\sum_{q={-}\frac{\hat{N}}{2}}^{\frac{\hat{N}}{2}{-}1}\!\!\mathcal{G}(q, k_i)  \!\!\sum_{\ell={-}\frac{\hat{M}}{2}}^{\frac{\hat{M}}{2}{-}1}  \!\!\mathcal{F}(\ell,l,l_i,k_i) \notag\\
    &~~~~~~~~~~~~~~~~~~\times X_{\textrm{DD}}\left[[k{-}k_i{+}q]_N,[l{-}|| l_i/\mu||{+}\ell]_M\right], 
\end{align} 
where $0<\hat{N} \leqslant N$ and $0<\hat{M} \leqslant M$.
Using this approximation will significantly decrease the computational complexity of signal detection for CP-OTFS-w-UCP-ECU, while only resulting in a slight degradation of the BER for a reasonably large set of $\hat{N}$ and $\hat{M}$ values~\cite{2018_TWC_Viterbo_OTFS_InterferenceCancellation}. 
The significance of this approximation will be illustrated through numerical results in Section~\ref{Sec:Num}-B.

Moreover,  we note that similar to that in~\eqref{Equ:ylk3_Approx}, the IORs for CP-OTFS-w-UCP and CP-OTFS-w-ECP in Corollaries 1 and 2 can also be approximated as 
\begin{align}\label{Equ:ylkSC_Approx}
    Y_{\textrm{DD}}[k,l] &= \sum_{i=1}^I h_i e^{j\frac{2 \pi (l-l_i)k_i}{NM}} \sum_{q=-\frac{\hat{N}}{2}}^{\frac{\hat{N}}{2}-1} \mathcal{G}(q, k_i) \notag\\
&~~~~~~~~~~~~~\times X_{\textrm{DD}}\left[[k-k_i+q]_N,[l-l_i]_M\right],
\end{align}
\begin{align}\label{Equ:ylkSC2_Approx}
    Y_{\textrm{DD}}[k,l] &= \sum_{i=1}^I h_i e^{j\frac{2 \pi (l-l_i)k_i}{NM}} \sum_{q=-\frac{\hat{N}}{2}}^{\frac{\hat{N}}{2}-1} \mathcal{G}_{\textrm{SC}}(q,k_i) \notag\\
&~~~~~~~~~~~~~\times X_{\textrm{DD}}[[k-k_i+q]_N,[l-l_i]_M].
\end{align} 
\end{remark}

\section{Channel Estimation}

\label{Sec:CE}

Accurate channel estimation (CE) is imperative to perform data detection for OTFS modulation. 
In this section, we propose an
embedded pilot-aided interference cancellation-based CE technique for CP-OTFS-w-UCP.

To facilitate channel estimation, we first rearrange the DD domain signal by adding a pilot symbol, and then several guard symbols surrounding the pilot symbols,  in a manner that effectively prevents interference between the pilot and data symbols in the DD domain~\cite{2019_TVT_YiHong_ChannelEstimationforOTFS}. Specifically, we let 
\begin{align}\label{gtx}
\!\!\!\!X_{\textrm{DD}}[k,l]&\!=\!\!\begin{cases}
x_p, & \!\!k=k_p,l=l_p,\\
0, & \!\!k_p{-}2\hat{k}_{\mathrm{max}}\leqslant k\leqslant k_p{+}2\hat{k}_{\mathrm{max}},\!\!\!\\
 & ~~~~l_p{-}l_{\mathrm{max}}\leqslant l\leqslant l_p{+}l_{\mathrm{max}},\!\!\!\\
X_{\textrm{DD,d}}[k,l], & \!\!\textrm{otherwise},
\end{cases}
\end{align}
where $x_p$ denotes the pilot symbol, $[k_p,l_p]$ denotes the arbitrary DD domain grid location for the pilot symbol, and $X_{\textrm{DD,d}}[k,l]$ denotes the data symbols at the $(k,l)$-th DD domain grid location.
Also, $\hat{k}_{\mathrm{max}}=k_{\mathrm{max}}+\hat{N}$ with
$k_{\mathrm{max}}=|\nu_{\mathrm{max}}|N T$ and $l_{\mathrm{max}}=\tau_{\mathrm{max}}M \Delta f$, where $\nu_{\mathrm{max}}$ and $\tau_{\mathrm{max}}$ are the maximum of the Doppler and the delay values of the propagation paths, respectively.

The threshold-based CE technique has been widely considered in the literature for estimating the DD domain channel~\cite{2019_TVT_YiHong_ChannelEstimationforOTFS}. In this method, the received DD domain symbols which carry the pilot power, $Y_{\mathrm{ch}}[l,k]=Y_{\textrm{DD}}[l+l_p,k+k_p-\hat{k}_{\mathrm{max}}]$, where $l\in \mathcal{L}_{\textrm{P}}=\{0, \cdots, l_{\mathrm{max}}\}$, $k\in \{0,\cdots,2\hat{k}_{\mathrm{max}}\}$, are directly used for CE while applying a positive CE threshold to avoid false detections of noise as propagation paths. Specifically, the channel coefficient associated with the Doppler tap $k$ and delay tap $l$ is estimated as
\begin{align}\label{XXX}
\hat{h}_{\mathrm{thr}}^{k,l}&=\begin{cases}
\frac{Y_{\mathrm{ch}}[l,k+\hat{k}_{\mathrm{max}}+1]}{x_p e^{j\frac{2 \pi l_p k}{NM}}}, & ~~~\textrm{if}~Y_{\mathrm{ch}}[l,k+\hat{k}_{\mathrm{max}}+1]\geqslant \mathcal{T}_{\mathrm{thr}},\\
0, & ~~~\textrm{otherwise},
\end{cases}
\end{align}
where $\mathcal{T}_{\mathrm{thr}}$ denotes the positive CE threshold using in the threshold-based CE technique.
It is noted that the threshold-based CE technique provides reasonable BER performance even with moderate pilot power when the channel response does not exhibit spreading effects along the Doppler (or/and delay) dimension. For instance, in the case of RCP-OTFS when propagation paths have on-grid Dopplers and on-grid delays, this method has been reported to be effective~\cite{2019_TVT_YiHong_ChannelEstimationforOTFS}.

However, in scenarios where channel response exhibits spreading effects along the Doppler (or/and delay) dimension, such as that of CP-OTFS-w-UCP which is the focus of this section\footnote{The CE for CP-OTFS-w-UCP-ECU, for which channel response exhibits spreading effects along both Doppler and delay dimensions will be discussed in our future works.},
the threshold-based CE technique would demand extremely high pilot power to accurately characterize the channel~\cite{2019_TVT_YiHong_ChannelEstimationforOTFS,2022_TWC_PAPRofOTFS,2022_TWC_Zhigiang_OffGridCEforOTFS}.
This is because the true channel coefficients and Dopplers of each propagation path are not determined by the threshold-based CE technique when spreading occurs along the Doppler (or/and delay) dimension. Instead, the channel coefficients of the perceived multiple taps that correspond to the actual propagation path are only estimated. 
As a result, extremely high pilot power may be needed to attain reasonable
BER performance for CP-OTFS-w-UCP when threshold-based CE technique is used.
While allocating high pilot power is theoretically possible, this can cause a high PAPR during practical implementation~\cite{2022_TWC_PAPRofOTFS,2022_TWC_Zhigiang_OffGridCEforOTFS}.

To overcome this challenge, we propose an interference cancellation-based CE technique for CP-OTFS-w-UCP. This technique characterizes the channel using the IOR derived in~\eqref{Equ:ylkSC_Approx} in Remark~\ref{Rem:GFApprox} and the received DD domain symbols which carry the pilot power, $Y_{\mathrm{ch}}[l,k]$. We estimate the complex channel coefficients and Doppler indices of the propagation paths associated with delay bins $l\in \mathcal{L}_{\textrm{P}}$ one after the other.

For a given delay bin, we assume there exists a maximum of $\bar{I}_l$ propagation paths, and then estimate the channel coefficients and Dopplers of those $ \bar{I}_l$ propagation paths.
We note that the channel coefficients and Dopplers of the $\bar{I}_l$ propagation path of the $l$th delay bin can be estimated by maximizing the log-likelihood function, or minimizing the Euclidean distance as 
\begin{align}\label{Equ:L1}
   \!\!\mathcal{L}\left({\boldsymbol \theta}_{l}/\mathbf{y}_{\mathrm{ch},l}\right)&{=}\left|\mathbf{y}_{\mathrm{ch},l}{-}\!\!\sum_{\iota=1}^{\bar{I}_{l}}h_{l,\iota} { \boldsymbol{\Psi}}_{l, \iota}x_p\right|^2,~~~~~\forall~ l\in\mathcal{L}_{\textrm{P}},
\end{align}
where $\mathbf{y}_{\mathrm{ch},l}=[Y_{\mathrm{ch}}[l,0], Y_{\mathrm{ch}}[l,1], \cdots, Y_{\mathrm{ch}}[l,2\hat{k}_{\mathrm{max}}]]^{\textrm{T}}$ and
${ \boldsymbol{\Psi}}_{l, \iota}= e^{j\frac{2 \pi l_p k_{l, \iota}}{NM}}\times [\mathcal{G}({-}\hat{k}_{\mathrm{max}},k_{l, \iota}), \mathcal{G}({-}\hat{k}_{\mathrm{max}}{+}1,$ $\cdots,  $ $\mathcal{G}(0,k_{l, \iota}),\cdots,\mathcal{G}(\hat{k}_{\mathrm{max}}{-}1,k_{l, \iota}),\mathcal{G}(\hat{k}_{\mathrm{max}},k_{l, \iota})]^{\textrm{T}}$  are  $(2\hat{k}_{\mathrm{max}}{+}1)\times 1$ vectors, $h_{l,\iota}$ and $k_{l,\iota}$ are the channel coefficient and the Doppler index of the $\iota$th propagation paths in the $l$th delay bin, respectively, and ${\boldsymbol \theta}_{l}={\boldsymbol \theta}^{\textrm{h}}_{l}\cup {\boldsymbol \theta}^{\textrm{k}}_{l}$ with ${\boldsymbol \theta}^{\textrm{h}}_{l}=\{h_{l, 1},h_{l, 2},\cdots,h_{l, \bar{I}_{l}}\}$, ${\boldsymbol \theta}^{\textrm{k}}_{l}=\{k_{l, 1},k_{l, 2},\cdots,k_{l, \bar{I}_{l}}\}$.
Mathematically, the maximum likelihood (ML)  estimator for the $l$th delay bin is written as 
\begin{align}\label{Equ:MLEst1}
   \hat{{\boldsymbol \theta}}_{l}&=\underset{\substack{{{\boldsymbol \theta}_{l}\in \mathbb{C}^{\bar{I}_{l}}\times \mathbb{Z}^{\bar{I}_{l}}}}}{\arg \min} \mathcal{L}\left({\boldsymbol \theta}_{l}/\mathbf{y}_{\mathrm{ch},l}\right).
\end{align}
A brute-force search in a $2\bar{I}_{l}$-dimensional
domain is necessary to find the solution to~\eqref{Equ:MLEst1}, which is unfeasible in general~\cite{2020_TWC_Lorenzo_OTFSforJSAC}. Thus, in the following, we propose a viable method to approximate the ML solution with low complexity.

We note that the $\mathcal{L}\left({\boldsymbol \theta}_{l}/\mathbf{y}_{\mathrm{ch},l}\right)$ in~\eqref{Equ:L1} is quadratic in $h_{l,\iota}$ for given $k_{l,\iota} $. We first differentiate $\mathcal{L}\left({\boldsymbol \theta}_{l}/\mathbf{y}_{\mathrm{ch},l}\right)$ in~\eqref{Equ:L1} with respect to (w.r.t.) $h_{l,\iota}$ and equate it to zero. In doing so, the minimization of~\eqref{Equ:MLEst1}  w.r.t. $h_{l,\iota}$ for fixed $k_{l,\iota} $ is readily
obtained as the solution to the linear system of equations 
\begin{align}\label{Equ:h1}
    \sum_{q=1}^{\bar{I}_{l}}h_{l,q} x^{\mathrm{H}}_p\boldsymbol{\Psi}^{\mathrm{H}}_{l, \iota}\boldsymbol{\Psi}_{l, q}x_p&=x^{\mathrm{H}}_p \boldsymbol{\Psi}^{\mathrm{H}}_{l, \iota}\mathbf{y}_{\mathrm{ch},l},~~~~~~~~\forall~ \iota\in\bar{\mathcal{I}}_l,
\end{align}
where $\bar{\mathcal{I}}_l=\{1,\cdots, {\bar{I}_{l}}\}$ and $()^{\mathrm{H}}$ denotes the Hermitian transpose operation. Thereafter, we expand~\eqref{Equ:L1} while using~\eqref{Equ:h1} and find that the minimization in~\eqref{Equ:MLEst1} 
reduces to maximizing the function 
\begin{align}\label{Equ:XX}
   &\!\!\!\!\!\mathcal{L}_2\left({\boldsymbol \theta}^{\textrm{k}}_{l}/\mathbf{y}_{\mathrm{ch},l}\right)=\sum_{\iota=1}^{\bar{I}_{l}}\underbrace{\frac{\left|x^{\mathrm{H}}_p {\boldsymbol{\Psi}}_{l,\iota}^{\mathrm{H}}\mathbf{y}_{\mathrm{ch},l}\right|^2}{\left|\boldsymbol{\Psi}_{l, \iota}x_p\right|^2}}_{\substack{\mathcal{L}_2^{\textrm{S}}}}\notag\\
   &~~~~~~~~~~~~~-\underbrace{\frac{\left(\sum_{q\neq\iota}h_{l,q} x^{\mathrm{H}}_p\boldsymbol{\Psi}^{\mathrm{H}}_{l, \iota}\boldsymbol{\Psi}_{l, q}x_p\right) \mathbf{y}^{\mathrm{H}}_{\mathrm{ch},l}{\boldsymbol{\Psi}}_{l,\iota}x_p }{\left|\boldsymbol{\Psi}_{l, \iota}x_p\right|^2}}_{\substack{\mathcal{L}_2^{\textrm{I}}}},\!\!\!\!\!
\end{align} 
where $\mathcal{L}_2^{\textrm{S}}$ and $\mathcal{L}_2^{\textrm{I}}$ denote the useful signal and the interference for the $\iota$th propagation path, respectively.
Thus, the ML estimator to identify ${\boldsymbol \theta}^{\textrm{k}}_{l}$ can be written as
\begin{align}\label{Equ:MLEst2}
   \hat{{\boldsymbol \theta}}^{\textrm{k}}_{l}&=\underset{\substack{{{\boldsymbol \theta}^{\textrm{k}}_{l}\in \mathbb{Z}}}}{\arg \max}~ \mathcal{L}_2\left({\boldsymbol \theta}^{\textrm{k}}_{l}/\mathbf{y}_{\mathrm{ch},l}\right).
\end{align}
Clearly, since $h_{l,\iota}$ $\forall \iota\in \bar{\mathcal{I}}_l$ are not known, it is impossible to directly maximize
$\mathcal{L}_2\left({\boldsymbol \theta}^{\textrm{k}}_{l}/\mathbf{y}_{\mathrm{ch},l}\right)$ in~\eqref{Equ:MLEst2} w.r.t. $k_{l,\iota}$. Furthermore, even for known $h_{l,\iota}$, the function $\mathcal{L}_2\left({\boldsymbol \theta}^{\textrm{k}}_{l}/\mathbf{y}_{\mathrm{ch},l}\right)$ is not separable w.r.t. $k_{l,\iota}$ for different values of $\iota$, due to the dependency of the interference terms $\mathcal{L}_2^{\textrm{I}}$ on all $k_{l,q}$,
for $q\neq \iota$.
Thus, we resort to an interference cancellation-based algorithm to determine $h_{l,\iota}$ and $k_{l,\iota}$, $\forall \iota\in \bar{\mathcal{I}}_l$, which is detailed as follows:

For notational convenience, consider the $\iota$th propagation path in a given delay bin to be the $\iota$th strongest propagation path in that delay bin.
In the first iteration of the algorithm, by assuming that there exists only one propagation path in the $l$th delay bin, we first estimate the Doppler index of the strongest propagation path in the $l$th delay bin, i.e., $\hat{{k}}_{l, 1}$, using \eqref{Equ:MLEst2}. We note that when it is considered that only one propagation path exists in the $l$th delay bin, the brute-force search in \eqref{Equ:MLEst2} becomes that in a one-dimensional integer domain, which is much simpler than that in~\eqref{Equ:MLEst1}. 
Thereafter, we estimate the channel coefficient of the strongest propagation path in the $l$th delay bin, i.e., $\hat{{h}}_{l, 1}$, using~\eqref{Equ:h1}.
Subsequently, we remove from $\mathbf{y}_{\mathrm{ch},l}$ the estimated contribution from the strongest propagation path in the $l$th delay bin. In doing so, we obtain the residual of $\mathbf{y}_{\mathrm{ch},l}$ after the first iteration, which we denote by  $\hat{\mathbf{y}}^{(1)}_{\mathrm{res},l}$.

\begin{algorithm}[t]

\caption{: Interference cancellation-based CE algorithm.
}
\begin{algorithmic}[1] \label{Alg:Alg1}
\STATE \textbf{Initialization}: Let $\hat{\mathbf{y}}_{\mathrm{res},l}^{(0)}=\mathbf{y}_{\mathrm{ch},l}$, $\forall ~l \in\mathcal{L}_{\textrm{P}}$.
\STATE \textbf{For}: $l=0, \cdots, l_{\mathrm{max}}$
\STATE ~~~\textbf{For}: $\iota=1,2,\cdots, \bar{I}_{l}$
   %
   \STATE ~~~~~~Solve the modified version of the ML estimator in~\eqref{Equ:MLEst2}  to obtain   $\hat{k}_{l, \iota}$:~~~~~~~~~~~~~~~~~~~~~~~~~~~~~~~~ \\ ~~~~~~~~~~~~$\hat{k}_{l, \iota}=\underset{\substack{{k_{l, \iota}\in \mathbb{Z}}}}{\arg \max}~ \mathcal{L}_2\left(k_{l, \iota}/\hat{\mathbf{y}}_{\mathrm{res},l}^{(\iota-1)}\right)$,  where \\ ~~~~~~~~~~~~~~~ $\mathcal{L}_2\left(k_{l, \iota}/\hat{\mathbf{y}}_{\mathrm{res},l}^{(\iota-1)}\right)=\frac{\left|x^{\mathrm{H}}_p {\boldsymbol{\Psi}}_{l,\iota}^{\mathrm{H}}\hat{\mathbf{y}}_{\mathrm{res},l}^{(\iota-1)}\right|^2}{\left|\boldsymbol{\Psi}_{l, \iota}x_p\right|^2}$.
     \STATE ~~~~~~~~~Solve the modified version of the system of equations in~\eqref{Equ:h1} to obtain   $\hat{h}_{l, \alpha},~\forall~ \alpha\in\{1,\cdots, \iota\}$: \\ ~~~~~~~~~~~~$\sum_{q=1}^{\iota}h_{l,q} x^{\mathrm{H}}_p\boldsymbol{\Psi}^{\mathrm{H}}_{l, \alpha}\boldsymbol{\Psi}_{l, q}x_p=x^{\mathrm{H}}_p \boldsymbol{\Psi}^{\mathrm{H}}_{l, \alpha}\mathbf{y}_{\mathrm{ch},l},~~\forall~ \alpha\in\{1,\cdots, \iota\}$.
  \STATE  ~~~~~~Update $\hat{\mathbf{y}}_{\mathrm{res},l}^{(\iota)}$:~~~~~~~~~~~~~~~~~~~~~~~~~~~~~~~~ \\ ~~~~~~~~~~~~$\hat{\mathbf{y}}_{\mathrm{res},l}^{(\iota)}=\mathbf{y}_{\mathrm{ch},l}-\sum_{\alpha=1}^{\iota}h_{l,\alpha} \boldsymbol{\Psi}_{l, \alpha}x_p$.
\STATE ~~~\textbf{End For}
\STATE  ~~~Update $\hat{h}_{l, \alpha},~\forall~ \alpha\in\{1,\cdots, \bar{I}_{l}\}$, using a positive CE threshold:~~~~~~~~~~~~~~~~~~ \\ ~~~~~~~~~$\hat{h}_{l,\alpha}=\begin{cases}
\hat{h}_{l,\alpha}, ~~~~~~~~\textrm{if}~~~~\hat{h}_{l,\alpha}\geqslant \mathcal{T},\\
0, ~~~~~~~~~~~\textrm{otherwise}.
\end{cases}
$
\STATE \textbf{End For}
\end{algorithmic}
\end{algorithm}

In the second iteration, we first estimate the Doppler index of the second strongest propagation path in the $l$th delay bin, i.e., $\hat{{k}}_{l, 2}$, using \eqref{Equ:MLEst2} while considering $\mathbf{y}_{\mathrm{ch},l}$ and $\bar{I}_{l}$ in~\eqref{Equ:XX} as $\hat{\mathbf{y}}^{(1)}_{\mathrm{res},l}$ and one, respectively. For this estimation, we again use a brute-force search in a one-dimensional integer domain.
Second, while considering $\mathbf{y}_{\mathrm{ch},l}$ and $\bar{I}_{l}$ in~\eqref{Equ:h1} as $\hat{\mathbf{y}}^{(2)}_{\mathrm{res},l}$ and two, respectively,  we estimate the channel coefficient of the second strongest propagation path in the $l$th delay bin, i.e., $\hat{{h}}_{l, 2}$, while updating the estimate for $\hat{{h}}_{l, 1}$.
Third, we determine the residual of $\mathbf{y}_{\mathrm{ch},l}$ after the second iteration, $\hat{\mathbf{y}}^{(2)}_{\mathrm{res},l}$, by removing from $\mathbf{y}_{\mathrm{ch},l}$ the estimated contribution from the first and the second strongest propagation paths in the $l$th delay bin.
This process is continued until Doppler index and channel coefficient of the $\bar{I}_{l}$th strongest propagation path in the $l$th delay bin is estimated using the corresponding residual of $\mathbf{y}_{\mathrm{ch},l}$, $\hat{\mathbf{y}}^{(\bar{I}_{l}-1)}_{\mathrm{res},l}$.
Finally, to avoid propagation paths being falsely detected as noise, a positive CE threshold, $\mathcal{T}$, is applied to the channel coefficient values.
The summary of the interference cancellation-based algorithm, along with the corresponding mathematical equations, is presented in Algorithm~1.

\vspace{5mm}
\section{Numerical Results}
\label{Sec:Num}

In this section, we present numerical results to highlight the considerations of this work. 
We adopt the Extended Vehicular A
(EVA) channel model and consider the carrier frequency to be $5~\mathrm{GHz}$, $\Delta f= 15~\mathrm{kHz}$, and user equipment (UE) speed to be $500~\mathrm{km/h}$~\cite{2018_TWC_Viterbo_OTFS_InterferenceCancellation}. 
We also consider $T_{\mathrm{reg}}^{\mathrm{cp}}=4.69 \mu \mathrm{s}> \tau_{\textrm{max}}=2.51 \mu \mathrm{s}$, $T_{\mathrm{long}}^{\mathrm{cp}}=5.2 \mu \mathrm{s}$, 4-QAM signaling,
 and the MP algorithm for signal detection~\cite{2018_TWC_Viterbo_OTFS_InterferenceCancellation}.
 Unless specified otherwise, the values for $M$ and $S$ are set to be $M\approx 0.6\times M'$ and $S=7$.

\subsection{Validation of Theorem 1 and Corollaries 1 and 2}

\begin{figure}[t]
\centering
\includegraphics[width=0.95\columnwidth]{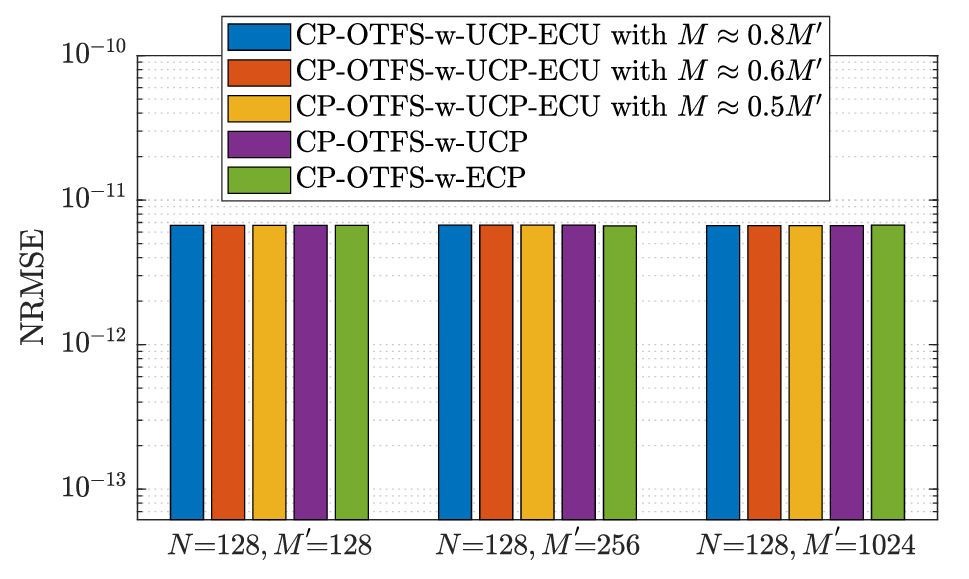}
\caption{\textcolor{black}{Illustration of the correctness of our derivations in Theorem 1 and Corollaries 1 and 2 using NRMSE for different $N$, $M$, and $M'$
values.}}\label{Fig:NumFigA} 
\end{figure}

In Fig.~\ref{Fig:NumFigA}, we first  validate our derivations in Theorem 1 and Corollaries 1 and 2 for different $N$, $M$, and $M'$
settings. To this end, we conduct simulations of CP-OTFS-w-UCP-ECU, CP-OTFS-w-UCP, and CP-OTFS-w-ECP
in a noiseless channel. We then compare the resulting received DD domain signals 
with those obtained from our derivations in Theorem 1 and Corollaries 1 and 2, using normalized root mean squared error (NRMSE). 
We observe that the NRMSE is negligibly small for CP-OTFS-w-UCP-ECU, CP-OTFS-w-UCP, and CP-OTFS-w-ECP for all $N$, $M$, and $M'$
settings. This validates our derivations in Theorem 1 and Corollaries 1 and 2.

\begin{figure}[!t]
\centering\subfloat[NRMSE associated with the approximation in~\eqref{Equ:ylk3_Approx} in Remark~\ref{Rem:GFApprox} for different $\hat{N}$ and $\hat{M}$ values.]{ \includegraphics[width=0.8\columnwidth]{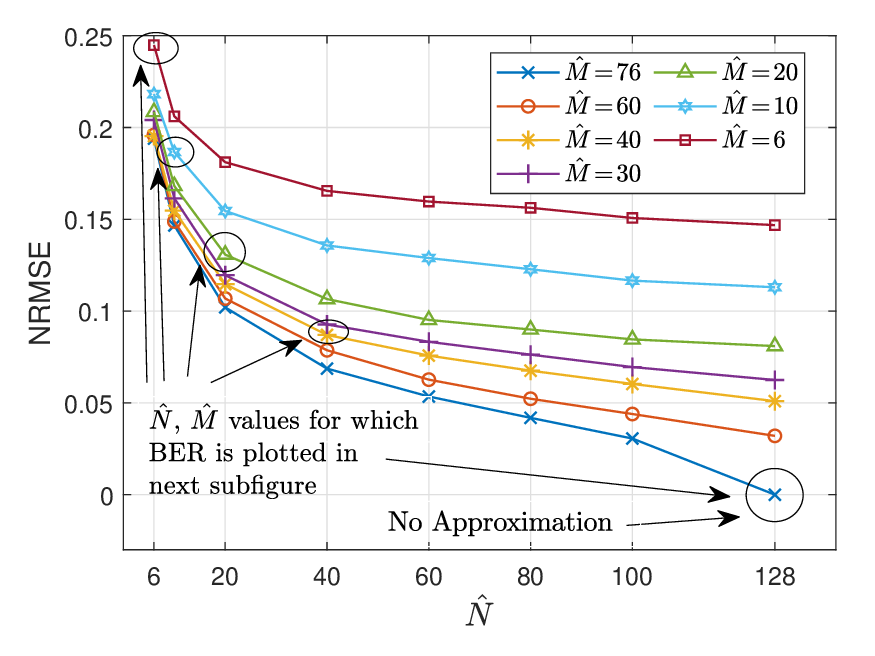}} \hspace{4 mm} \subfloat[BER versus $\mathrm{SNR}_{\mathrm{d}}$ when signal detection is performed using the IOR derived in (i) Theorem 1 and (ii) the approximation in~\eqref{Equ:ylk3_Approx} in Remark~\ref{Rem:GFApprox}.]{\includegraphics[width=0.8\columnwidth]{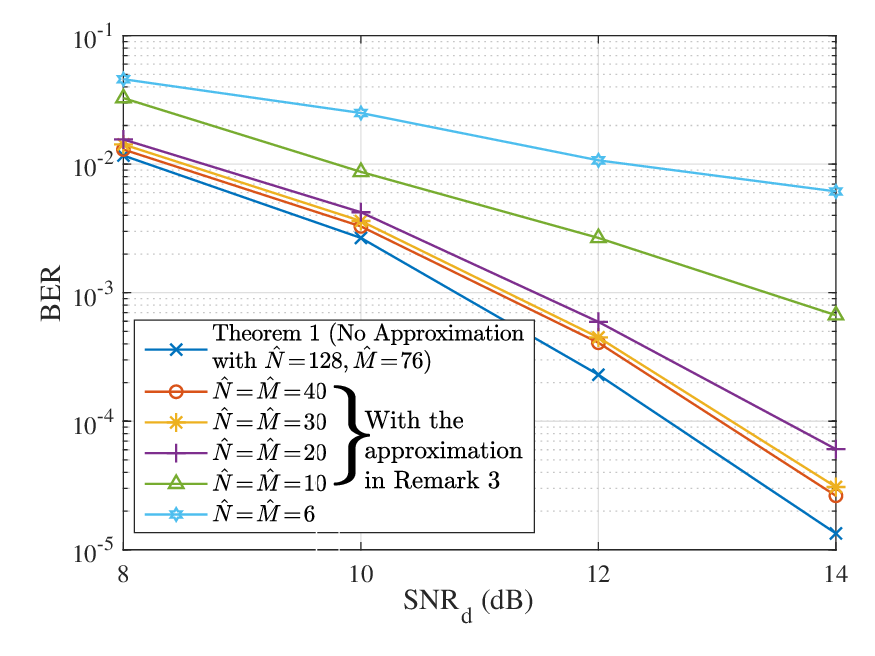}}
  \hspace{2 mm}
  \subfloat[Percentage reduction in computational complexity achieved when using the approximation given in~\eqref{Equ:ylk3_Approx} in Remark~\ref{Rem:GFApprox}.]{ \includegraphics[width=0.8\columnwidth]{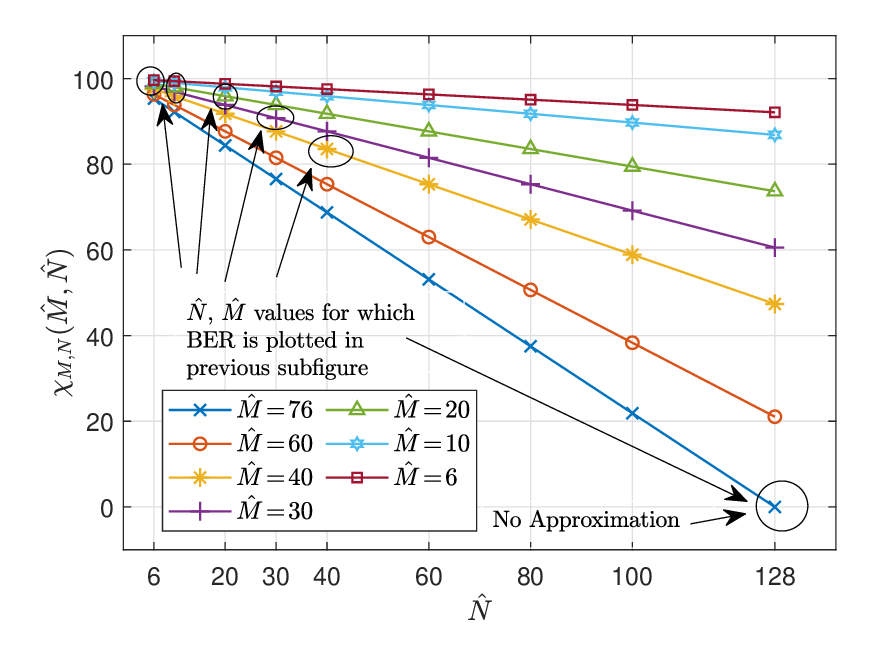}}
  \caption{Investigation of the approximation introduced in~\eqref{Equ:ylk3_Approx} in Remark~\ref{Rem:GFApprox} for the IOR of CP-OTFS-w-UCP-ECU.}\label{Fig:NumFigB}
\end{figure}

\subsection{Approximate IORs Introduced in Remark~\ref{Rem:GFApprox}}

In Fig.~\ref{Fig:NumFigB},  we delve into the approximation described in~\eqref{Equ:ylk3_Approx} in Remark~\ref{Rem:GFApprox} for the IOR of CP-OTFS-w-UCP-ECU. To this end, we first conduct simulations of CP-OTFS-w-UCP-ECU  in a noiseless channel. Then,  in Fig.~\ref{Fig:NumFigB} (a), we compare the resulting received DD domain signals
with those obtained from the approximate IOR in Remark~\ref{Rem:GFApprox}  while choosing different $\hat{N}$ and $\hat{M}$ values for the approximation.
Second,
we simulate CP-OTFS-w-UCP-ECU  in a noisy channel and then perform signal detection  while using the IOR derived in (i) Theorem 1 and (ii) the approximation in Remark~\ref{Rem:GFApprox}. The resulting  BERs are plotted versus the signal-to-noise-ratio (SNR) of data symbols, $\mathrm{SNR}_{\mathrm{d}}$, for different $\hat{N}$ and $\hat{M}$ values  in Fig.~\ref{Fig:NumFigB}(b).
Finally, in Fig.~\ref{Fig:NumFigB}(c) we plot  the percentage reduction in computational complexity achieved when using the approximate IOR described in Remark~\ref{Rem:GFApprox} instead of the IOR derived from Theorem~1, $\boldsymbol{\chi}_{M,N}(\hat{N},\hat{M})$. For the MP algorithm based signal detection, $\boldsymbol{\chi}_{M,N}(\hat{N},\hat{M})$ is given by $\boldsymbol{\chi}_{M,N}(\hat{N},\hat{M})=\frac{NM-\hat{N}\hat{M}}{NM}$.
Considering the high computational complexity of the MP signal detection algorithm, in Fig.~\ref{Fig:NumFigB}, as well as in all the proceeding figures, we let $N=M'=128$ and $M=76$. 

From Fig.~\ref{Fig:NumFigB}(a) we observe that the NRMSE associated with the approximate IOR in Remark~\ref{Rem:GFApprox} increases significantly as $\hat{N}$ and $\hat{M}$ decreases. However, from Fig.~\ref{Fig:NumFigB}(b) we observe that although the resulting BER degradation is very high for low $\hat{N}$ and $\hat{M}$ values, the BER degradation is marginal for certain high values of $\hat{N}$ and $\hat{M}$, e.g., $\hat{N},\hat{M}\geqslant 30$ when $N=M'=128$. The marginal degradation in BER is noteworthy, especially considering that the resulting reduction in computational complexity is significant,  as can be observed from Fig.~\ref{Fig:NumFigB}(c).
For example, when $\hat{N}=\hat{M}=30$, only a $0.5~\mathrm{dB}$ power loss occurs at the BER of $10^{-3}$, while the resulting reduction in computational complexity during signal detection is $90.75\%$.
These findings show that it may be beneficial to utilize the approximate IOR described in Remark~\ref{Rem:GFApprox} for signal detection for CP-OTFS-w-UCP-ECU, provided that suitable values of $\hat{N}$ and $\hat{M}$ are chosen for the approximation, such as $\hat{N}=\hat{M}=30$ when $N=M'=128$~\cite{2018_TWC_Viterbo_OTFS_InterferenceCancellation}.\footnote{\textcolor{black}{We note that the specific values of $\hat{M}$ and $\hat{N}$ that can simplify the IOR would vary depending on many channel and the system parameters, including the values of $I$, the maximum of the Doppler and the delay values of the propagation paths ($\nu_{\mathrm{max}}$ and $\tau_{\mathrm{max}}$), $M$, $N$, $M'$, $S$, $T_{\mathrm{long}}^{\mathrm{cp}}$, and $T_{\mathrm{reg}}^{\mathrm{cp}}$. This necessitates devising an approach that can identify suitable values for $\hat{M}$ and $\hat{N}$ for any given set of channel and system parameters.  However, devising such an approach is beyond the scope of this work, but will be considered in our future works.}}

\begin{figure}[t]
\centering
\includegraphics[width=0.99\columnwidth]{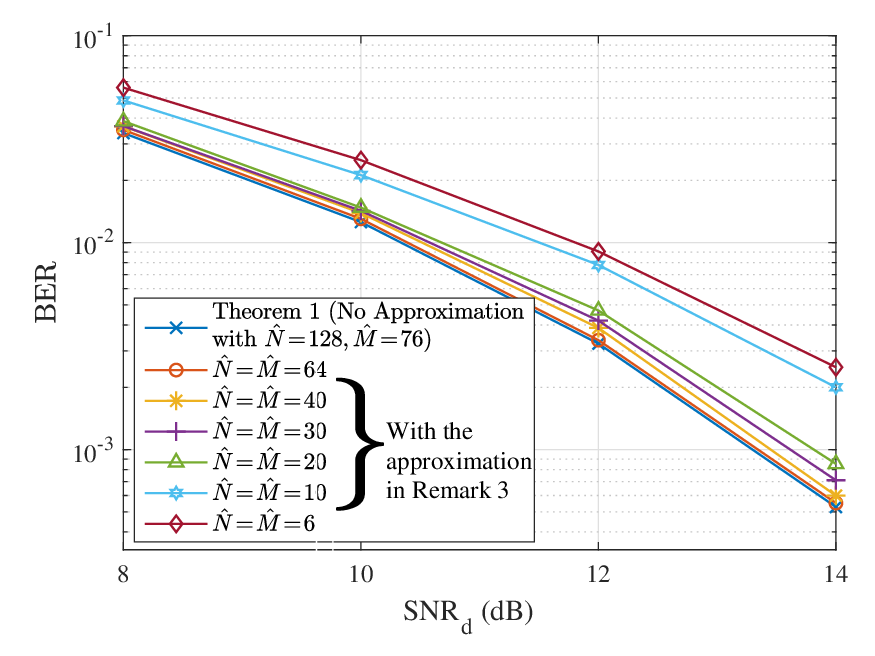}
\caption{Investigation of the approximation introduced in ~\eqref{Equ:ylkSC_Approx} in Remark~\ref{Rem:GFApprox} for the IOR of CP-OTFS-w-UCP.}\label{Fig:NumFigC} 
\end{figure}

Next, similar to Fig.~\ref{Fig:NumFigB} where we investigate the approximation associated with  CP-OTFS-w-UCP-ECU, in Fig.~\ref{Fig:NumFigC}  we investigate the approximation described in~\eqref{Equ:ylkSC_Approx} in Remark~\ref{Rem:GFApprox} for the IOR of CP-OTFS-w-UCP.
To this end, we simulate CP-OTFS-w-UCP in a noisy channel and perform signal detection while using the IOR derived in (i) Corollary 1 and (ii) the approximation in Remark~\ref{Rem:GFApprox}. The resulting BERs are plotted versus $\mathrm{SNR}_{\mathrm{d}}$ for different $\hat{N}$ and $\hat{M}$ values.
Similar to that in Fig.~\ref{Fig:NumFigB}(b) for CP-OTFS-w-UCP-ECU, we observe that the BER degradation for CP-OTFS-w-UCP is marginal for certain $\hat{N}$, e.g., $\hat{N}\geqslant 20$ when $N=128$, and thus conclude that it is beneficial to utilize the approximate IOR described in Remark~\ref{Rem:GFApprox} for signal detection for CP-OTFS-w-UCP-ECU, provided that a suitable value of $\hat{N}$ is chosen for the approximation, such as $\hat{N}=20$ when $N=M'=128$.

\subsection{The Impact CPs of Unequal Lengths}

\begin{figure}[!t]
\centering\subfloat[$S=7$. ]{ \includegraphics[width=0.85\columnwidth]{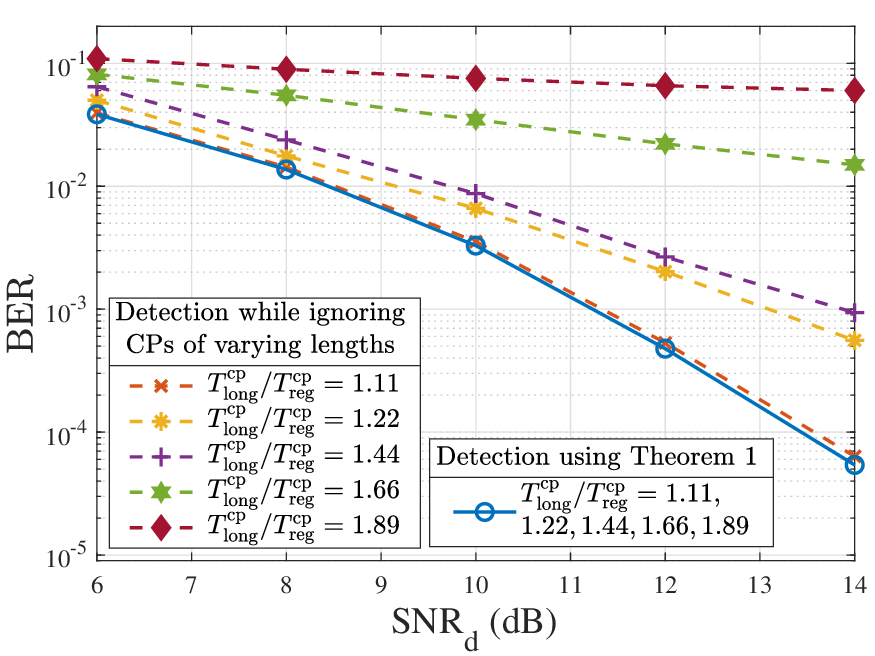}} \hspace{4 mm} \subfloat[$S=14$.]{\includegraphics[width=0.85\columnwidth]{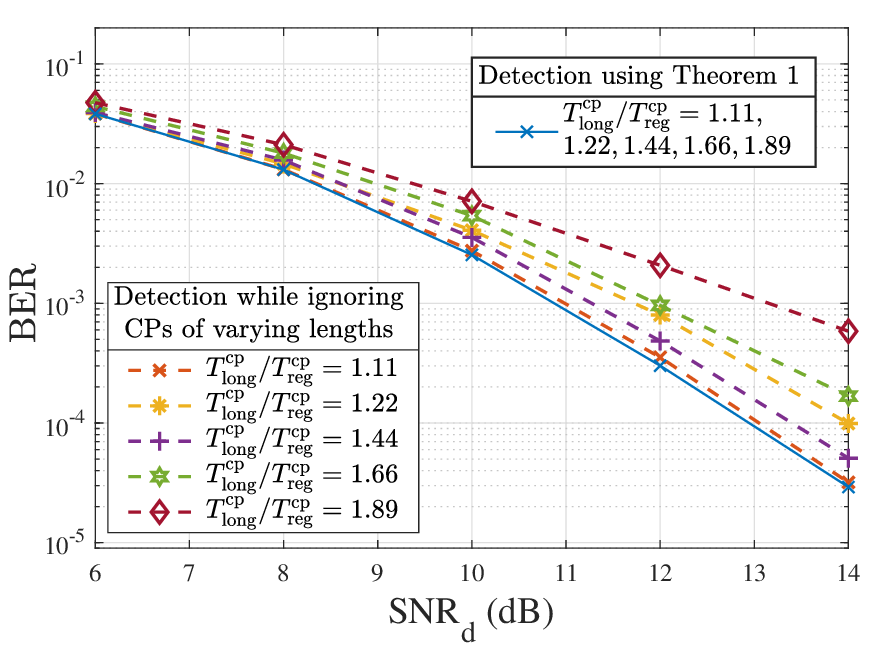}}
 \hspace{2 mm}
  \subfloat[$S=28$.]{ \includegraphics[width=0.85\columnwidth]{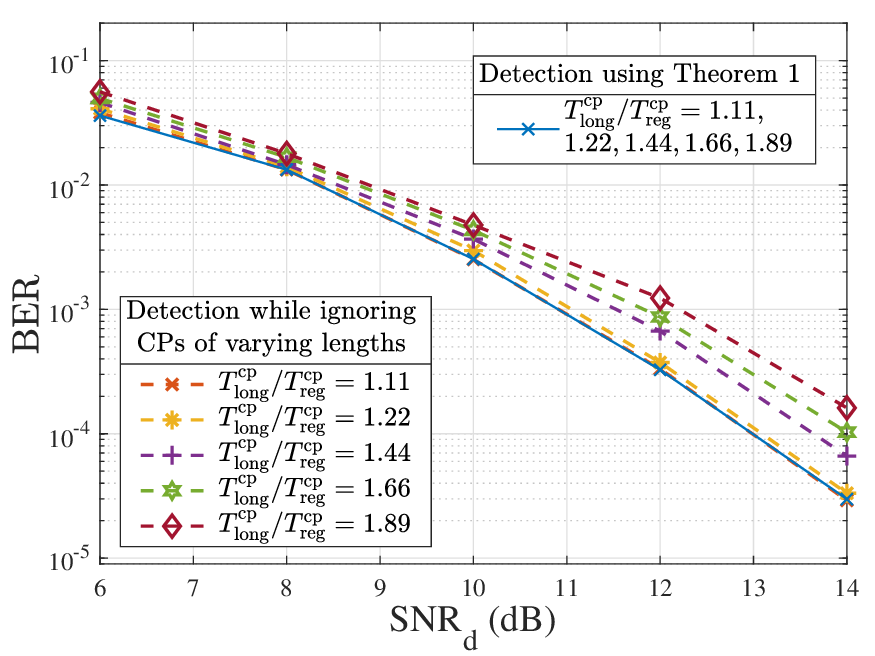}}
  \caption{Illustration of the  impact of not ignoring the unequal lengths of CPs.}\label{Fig:NumFigD}
\end{figure}

In Fig.~\ref{Fig:NumFigD}, we illustrate the importance of explicitly accounting for the unequal lengths of the CPs when OTFS coexists with OFDM systems. To this end, we simulated CP-OTFS-w-UCP-ECU in a noisy channel for different values of the ratio $T_{\mathrm{long}}^{\mathrm{cp}}/T_{\mathrm{reg}}^{\mathrm{cp}}$ and the number of OFDM signals carried by a time window ($S$). We then performed signal detection while using the IOR derived in~\eqref{Equ:ylk3_Approx} in Remark~\ref{Rem:GFApprox}.\footnote{In each figure, a single curve is plotted for different values of $T_{\mathrm{long}}^{\mathrm{cp}}/T_{\mathrm{reg}}^{\mathrm{cp}}$ when signal detection is performed using the IOR derived in Theorem 1. This is because for different values of $T_{\mathrm{long}}^{\mathrm{cp}}/T_{\mathrm{reg}}^{\mathrm{cp}}$, similar BER values are obtained when simulations are carried out for a sufficiently large number of simulation trials.} 
For comparison, we perform signal detection for CP-OTFS-w-UCP-ECU while using the IOR in which the impact of unequal lengths of CPs is ignored.

\textcolor{black}{First, it can be observed that when $T_{\mathrm{long}}^{\mathrm{cp}}/T_{\mathrm{reg}}^{\mathrm{cp}}$ is high and $S$ is small, e.g., $T_{\mathrm{long}}^{\mathrm{cp}}/T_{\mathrm{reg}}^{\mathrm{cp}}=1.4,1.6, 1.8$ with $S=7$ as in Fig.~\ref{Fig:NumFigD}(a), BER deterioration due to ignoring the impact of unequal lengths of CPs is very high. 
We clarify that when $T_{\mathrm{long}}^{\mathrm{cp}}/T_{\mathrm{reg}}^{\mathrm{cp}}$ is high, the influence of long CPs in the OTFS signal is high. Also, when $S$ is small, the number of OFDM signals with long CPs within the OTFS signal is high, thereby increasing the influence of long CPs in the OTFS signal. These lead to the BER deterioration. 
These BER deteriorations show that it may not be acceptable to ignore the impact of unequal lengths of CPs during signal detection when $T_{\mathrm{long}}^{\mathrm{cp}}/T_{\mathrm{reg}}^{\mathrm{cp}}$ is very high and $S$ is small.}

Second, by comparing the curves in a single subfigure, we observe that the BER deterioration decreases when $T_{\mathrm{long}}^{\mathrm{cp}}/T_{\mathrm{reg}}^{\mathrm{cp}}$ decreases, and this results in negligibly small BER deterioration for low $T_{\mathrm{long}}^{\mathrm{cp}}/T_{\mathrm{reg}}^{\mathrm{cp}}$ for all values of $S$, e.g. $T_{\mathrm{long}}^{\mathrm{cp}}/T_{\mathrm{reg}}^{\mathrm{cp}}=1.1$ with $S=7,14,28$. Third, by comparing the curves across subfigures, we observe that the BER deterioration decreases when $S$ increases. 
The second and third observations show that when $T_{\mathrm{long}}^{\mathrm{cp}}/T_{\mathrm{reg}}^{\mathrm{cp}}$ is low and/or $S$ is very high, 
it may be reasonable to perform signal detection while using the IOR in which the impact of unequal lengths of CPs is ignored. 
We clarify that this insight was made possible only because of our analysis on CP-OTFS with CPs of unequal lengths, thereby further highlighting the significance of this work.

\subsection{The Proposed CE Technique }

\subsubsection{BER Analysis}

\begin{figure}[t]
\centering
\includegraphics[width=0.99\columnwidth]{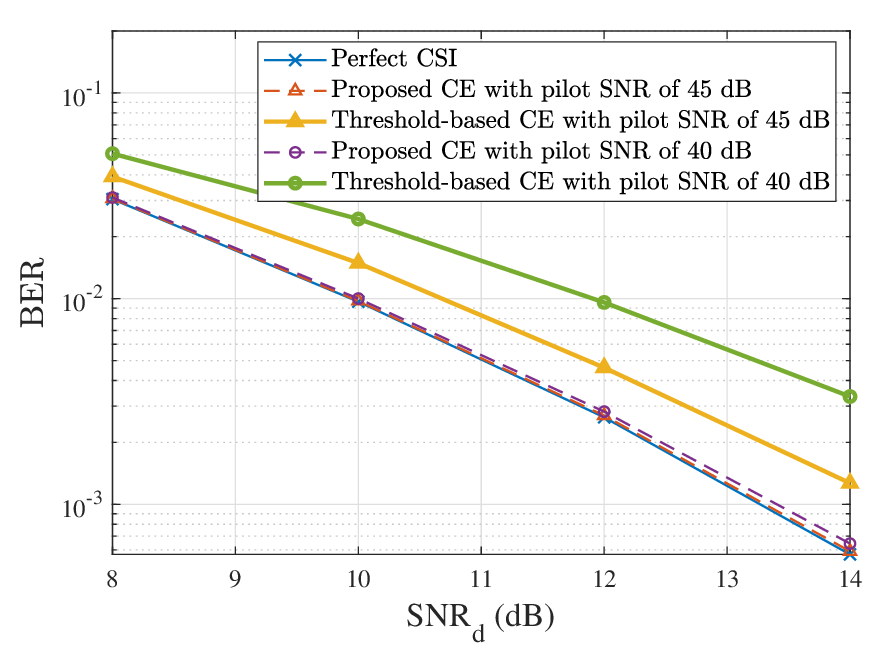}
\caption{\textcolor{black}{Illustration of BER versus $\mathrm{SNR}_{\mathrm{d}}$ for different CE techniques.}}\label{Fig:NumFigE} 
\end{figure}

To demonstrate the significance of the CE technique proposed in Section~\ref{Sec:CE}, in Fig.~\ref{Fig:NumFigE}, we plot the BER for CP-OTFS-w-UCP versus $\mathrm{SNR}_{\mathrm{d}}$ when the channel is estimated using the proposed CE technique. For comparison, we plot (i) the BER with perfect CSI and (ii) the BER when the channel is estimated using the state-of-the-art threshold-based CE technique~\cite{2019_TVT_YiHong_ChannelEstimationforOTFS}. 
Moreover, in Fig.~\ref{Fig:NumFigE}, we set $\bar{I}_l=5$, $\hat{N}=\hat{M}=20$, and $\mathcal{T}=3/\sqrt{\mathrm{SNR}_{\mathrm{p}}}$, where $\mathrm{SNR}_{\mathrm{p}}$ denotes the SNR of pilot symbols~\cite{2019_TVT_YiHong_ChannelEstimationforOTFS,2018_TWC_Viterbo_OTFS_InterferenceCancellation,2017_WCNC_OTFS_Haddani,2022_TWC_JH_ODDM}.
We first observe that for all $\mathrm{SNR}_{\mathrm{d}}$ values, the proposed CE technique outperforms the threshold-based CE technique. 
We also observe that the BER of the proposed CE technique approaches to that of perfect CSI. These observations show the significance of our proposed CE technique for OTFS when the channel response exhibits spreading effects.
\textcolor{black}{Moreover, we observe that although the BER of the threshold-based CE technique deteriorates significantly as the pilot power decreases, the deterioration in the BER for our proposed CE technique is marginal. 
Due to this, when the pilot SNR is low, a significant SNR improvement is attained by the proposed CE technique in comparison to the threshold-based CE technique. In particular, at the BER of $10^{-2}$, the SNR improvement is $2~\textrm{dB}$ for the pilot SNR of $40~\textrm{dB}$. 
This shows the significance of our proposed CE technique.}

\begin{figure}[t]
\centering
\includegraphics[width=0.95\columnwidth]{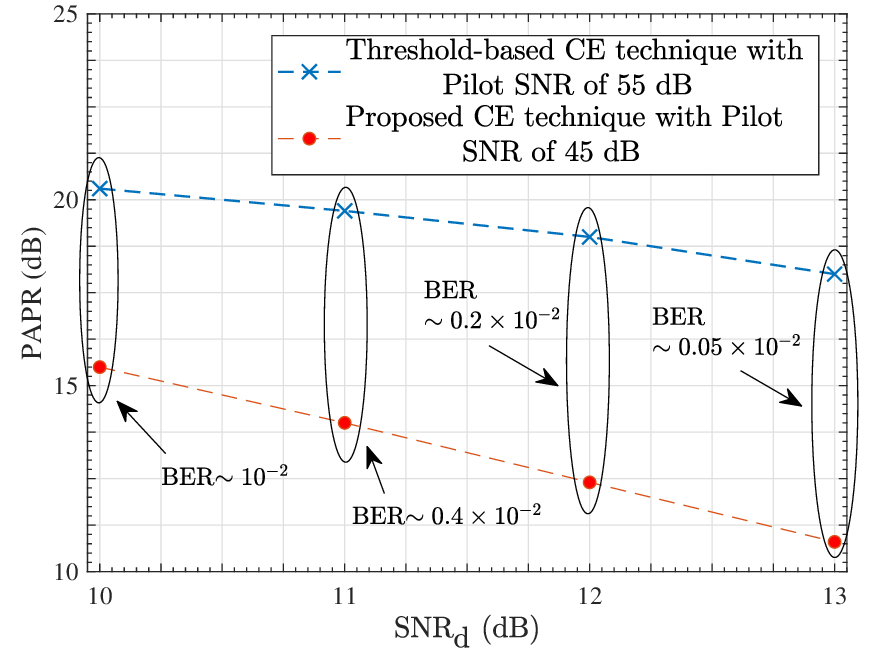}
\caption{\textcolor{black}{Illustration of the PAPR improvement brought by our proposed CE technique.}}\label{Fig:NumFigH} 
\end{figure}

\subsubsection{PAPR Analysis}
\textcolor{black}{In Fig.~\ref{Fig:NumFigH}, we examine the PAPR improvement brought by our proposed CE technique. 
To this end, we plot the PAPRs of the transmitted signals with different pilot energies that would lead to comparable BERs when the threshold-based CE technique~\cite{2019_TVT_YiHong_ChannelEstimationforOTFS} and our proposed CE techniques are utilized.
Based on the PAPR values, it is evident that our proposed CE technique can achieve a PAPR improvement, especially when the SNR of data symbols, $\mathrm{SNR}_{\mathrm{d}}$, is high.
}

\begin{figure}[t]
\centering
\includegraphics[width=0.95\columnwidth]{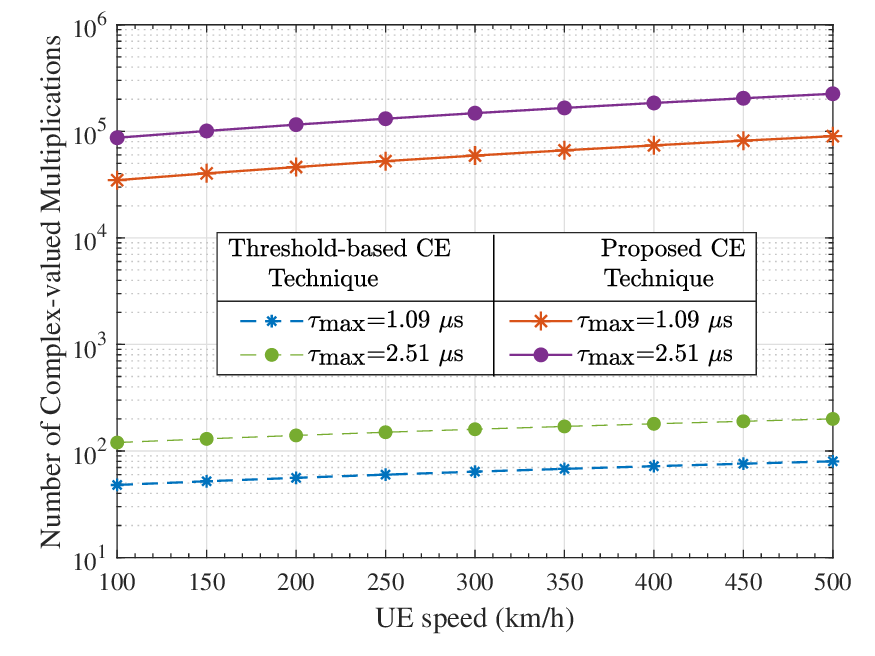}
\caption{\textcolor{black}{Illustration of the computational complexity our proposed CE technique.}}\label{Fig:NumFigI} 
\end{figure}

\subsubsection{\textcolor{black}{Complexity Analysis}}

\textcolor{black}{
Finally, in Fig.~\ref{Fig:NumFigI}, we examine the computational complexity of the proposed CE technique in comparison with that of the threshold-based CE technique.
To this end, we plot the number of complex-valued multiplications that dominates the computational complexity of both CE techniques for different  channel parameters~\cite{2022_TCOM_LowComplexityReceiverforOTFS}.\footnote{\textcolor{black}{Note that the EVA model is slightly modified to obtain the channel with maximum delay values of the propagation paths, $\tau_{\mathrm{max}}$ of $1.09~\mu$s.}} 
We note that the computational complexity of the threshold-based CE technique~\cite{2019_TVT_YiHong_ChannelEstimationforOTFS} is primarily dominated by the number of complex-valued multiplications, which is on the order of $\hat{k}_{\mathrm{max}}l_{\mathrm{max}}$.
As for the computational complexity of our proposed CE technique, it is primarily dominated by the operations involved in solving (i) the ML estimator given in line 4 of Algorithm 1 and (ii) the system of linear equations given in line 5 of Algorithm 1.
In particular, the complexity of solving the ML estimators is dominated by the number of complex-valued multiplications and complex-value additions, which are on the order of $\bar{I}^2_{l}\hat{k}_{\mathrm{max}}^2l_{\mathrm{max}}$, while the complexity of solving the system of linear equations is dominated by the number of complex-valued multiplications and complex-value additions, which are on the order of $\bar{I}^3_{l}\hat{k}_{\mathrm{max}}l_{\mathrm{max}}$~\cite{2022_TWC_Zhigiang_OffGridCEforOTFS}. 
}

\textcolor{black}{
As can be observed from Fig.~\ref{Fig:NumFigI}, the number of complex-valued multiplications associated with our proposed CE technique are much higher compared to those of the threshold-based CE technique. Based on this, it is apparent that the  improvement in CE accuracy that can be provided by our proposed CE technique comes at the expense of increased computational complexity compared  to the threshold-based CE technique.}

\section{Conclusion}

In this work, we investigated the coexistence of OTFS with current 4G/5G communication systems that use OFDM waveforms. We first derived the IOR of OTFS in coexisting systems while considering unequal lengths of CPs and ECU. We showed analytically that the inclusion of multiple CPs to the OTFS signal and ECU results in the channel response to exhibit spreading effects/leakage along the Doppler and delay dimensions, respectively. Consequently, we showed that the effective sampled DD domain channel model for OTFS in coexisting systems may exhibit reduced sparsity. 
Thereafter, we proposed an embedded pilot-aided interference cancellation-based CE technique for OTFS in coexisting systems that leverages the derived IOR for accurate channel characterization.  Using numerical results, we first validated our analysis. We then showed (i) that ignoring the impact of unequal lengths of CPs during signal detection can degrade the BER performance of  OTFS in coexisting systems, and (ii) the significance of our proposed CE technique.

\appendices

\section{Proof of Theorem~\ref{Thr:IORCPOTFS-VCP-ECU}}\label{App:IORCPOTFS-VCP-ECU}

\begin{figure*}[!t]
\normalsize  
\begin{align}\label{Equ:G2}
   &\mathcal{\tilde{G}}(k,\bar{k},k_i)
   =\frac{1}{N}\Big(\sum _{n=0}^{S-1} e^{-j\frac{2 \pi }{N}(n(k-\bar{k}-k_i(1+\psi^{\mathrm{reg}})))}+\sum _{n=S}^{2S-1} e^{-j\frac{2 \pi }{N}(n(k-\bar{k}-k_i(1+\psi^{\mathrm{reg}}))-k_i\psi^{\mathrm{ext}})}+\cdots\cdots\notag\\[-6pt]
   &~~~~~~~~~~~~~~~~~~~~~~~~+\sum _{n=N-S}^{\omega_{\textrm{f}} S-1} e^{-j\frac{2 \pi }{N}(n(k-\bar{k}-k_i(1+\psi^{\mathrm{reg}}))-(\omega_{\textrm{f}}-1)k_i\psi^{\mathrm{ext}})}-\sum _{n=N}^{\omega_{\textrm{f}} S-1} e^{-j\frac{2 \pi }{N}(n(k-\bar{k}-k_i(1+\psi^{\mathrm{reg}}))-(\omega_{\textrm{f}}-1)k_i\psi^{\mathrm{ext}})}\Big) \notag\\[-5pt]
   &=\underbrace{\frac{1}{N}\sum _{\beta=0}^{\omega_{\textrm{f}}-1} \sum _{\hat{n}=0}^{S-1} e^{-j\frac{2 \pi }{N}((\hat{n}+\beta S)(k-\bar{k}-k_i(1+\psi^{\mathrm{reg}}))-\beta k_i\psi^{\mathrm{ext}})}}_{\substack{\mathrm{T}^{\mathcal{G}}_{1}}} -\underbrace{\frac{1}{N}\sum _{\hat{n}=\omega_{\textrm{m}}}^{S-1} e^{-j\frac{2 \pi }{N}((\hat{n}+(\omega_{\textrm{f}}-1) S)(k-\bar{k}-k_i(1+\psi^{\mathrm{reg}}))-(\omega_{\textrm{f}}-1) k_i\psi^{\mathrm{ext}})}}_{\substack{\mathrm{T}^{\mathcal{G}}_{2}}}.
\end{align} 
\begin{align}\label{Equ:G3}
   \mathrm{T}^{\mathcal{G}}_{1}&=\frac{1}{N}\sum _{\beta=0}^{\omega_{\textrm{f}}-1} e^{-j\frac{2 \pi S\beta}{N}(k-\bar{k}-k_i(1+\psi^{\mathrm{reg}}+\frac{\psi^{\mathrm{ext}}}{S}))} \sum _{\hat{n}=0}^{S-1} e^{-j\frac{2 \pi \hat{n} }{N}(k-\bar{k}-k_i(1+\psi^{\mathrm{reg}}))}\notag\\
   &=\begin{cases}
\frac{\omega_{\textrm{f}} S}{N}\delta(-k+\bar{k}),& k_i=0,\\
\frac{e^{-j\frac{2\pi SP}{N} (k-\bar{k}-k_i(1+\psi^{\mathrm{reg}}+\frac{\psi^{\mathrm{ext}}}{S}))}-1}{e^{-j\frac{2 \pi S}{N}(k-\bar{k}-k_i(1+\psi^{\mathrm{reg}}+\frac{\psi^{\mathrm{ext}}}{S}))}-1}\frac{e^{-j\frac{2 \pi S}{N}(k-\bar{k}-k_i(1+\psi^{\mathrm{reg}}))}-1}{N e^{-j\frac{2 \pi }{N}(k-\bar{k}-k_i(1+\psi^{\mathrm{reg}}))}-N},& \textrm{elsewhere}.
\end{cases}\\
   \mathrm{T}^{\mathcal{G}}_{2}&=\frac{1}{N}e^{-j\frac{2 \pi S(\omega_{\textrm{f}}-1)}{N}(k-\bar{k}-k_i(1+\psi^{\mathrm{reg}}+\frac{\psi^{\mathrm{ext}}}{S}))} \sum _{\hat{n}=\omega_{\textrm{m}}}^{S-1} e^{-j\frac{2 \pi \hat{n} }{N}(k-\bar{k}-k_i(1+\psi^{\mathrm{reg}}))} \notag\\[-5pt]
   &=\begin{cases}
\frac{S-\omega_{\textrm{m}}}{N}\delta(-k+\bar{k}),& k_i=0,\\
e^{-j\frac{2\pi S(\omega_{\textrm{f}}-1)}{N} (k-\bar{k}-k_i(1+\psi^{\mathrm{reg}}+\frac{\psi^{\mathrm{ext}}}{S}))}\frac{e^{-j\frac{2 \pi S}{N}(k-\bar{k}-k_i(1+\psi^{\mathrm{reg}}))}-e^{-j\frac{2 \pi \omega_{\textrm{m}}}{N}(k-\bar{k}-k_i(1+\psi^{\mathrm{reg}}))}}{N e^{-j\frac{2 \pi }{N}(k-\bar{k}-k_i(1+\psi^{\mathrm{reg}}))}-N},& \textrm{elsewhere}.
\end{cases} \label{Equ:G4}
\end{align}
\begin{subequations}
\begin{align}
   \mathcal{\tilde{F}}(l,\bar{l},l_i,k_i)&=\frac{1}{MM'}\Big(\sum _{s=0}^{M'{-}1} e^{j\frac{2 \pi s k_i }{NM'}}\sum _{m=0}^{\frac{M}{2}{-}1} \zeta_1^m\sum _{\bar{m}=0}^{\frac{M'}{2}{-}1} \zeta_2^{\bar{m}}+\zeta_3\sum _{s=0}^{M'{-}1} \zeta_4^{-1}e^{j\frac{2 \pi s k_i }{NM'}}\sum _{m=0}^{\frac{M}{2}{-}1} \zeta_1^m\sum _{\bar{m}=\frac{M'}{2}}^{M{-}1} \zeta_2^{\bar{m}}\notag\\
   &~~~~~~~~~~~+\zeta_3\sum _{s=0}^{M'{-}1} e^{j\frac{2 \pi s k_i }{NM'}}\sum _{m=\frac{M}{2}}^{M{-}1} \zeta_1^m\sum _{\bar{m}=0}^{\frac{M'}{2}{-}1} \zeta_2^{\bar{m}}+\sum _{s=0}^{M'{-}1} \zeta_4e^{j\frac{2 \pi s k_i }{NM'}}\sum _{m=\frac{M}{2}}^{M{-}1} \zeta_1^m\sum _{\bar{m}=\frac{M}{2}}^{M'{-}1} \zeta_2^{\bar{m}}\Big)\\
   &=\begin{cases}
\delta(-l+\bar{l}+\frac{l_i}{\mu}),& k_i=[l_i]_\mu=0,\\
\frac{\zeta_5^{\frac{M}{2}}-1}{M(\zeta_5-1)}(1+\zeta_3\zeta_5^{\frac{M}{2}})
,& k_i=0, [l_i]_\mu \neq 0,\\
\sum_{s=0}^{M'-1}e^{j\frac{2\pi sk_i}{NM'}}\frac{\zeta_1^{\frac{M}{2}}-1}{M'(\zeta_1-1)}\frac{\zeta_2^{\frac{M}{2}}-1}{M(\zeta_2-1)}(1+\zeta_4\zeta_1^{\frac{M}{2}})(1+\zeta_3\zeta_4^{-1}\zeta_2^{\frac{M}{2}}),& \textrm{elsewhere}.
\end{cases} 
\end{align}\label{Equ:F3}
\end{subequations}

\hrulefill
\end{figure*}

We first start the proof of Theorem 1 by simplifying $\mathcal{\tilde{G}}(k,\bar{k},k_i)$ given in~\eqref{Equ:G}.
We note that for all 5G NR numerologies, the number of OFDM signals carried by a time window ($S$) is an integer multiple of seven (see Table~\ref{tab:OFDMNum})~\cite{3GPPStandOFDM}. On the other hand, to enable easier hardware implementation of the ISFFT operation in~\eqref{Equ:Xtf} and SFFT operation in~\eqref{Equ:ylk1}, $N$ may be set to be a power of two (refer Remark~\ref{Rem:SelectionofMMprim}). Due to these factors, $S$ may not necessarily be an integer multiple of the $N$, i.e., $[N]_S\neq 0$. Considering this, we expand the $\mathcal{\tilde{G}}(k,\bar{k},k_i)$ in~\eqref{Equ:G} to obtain~\eqref{Equ:G2}.
The expression~\eqref{Equ:G2}, along with~\eqref{Equ:G3},~\eqref{Equ:G4},  and~\eqref{Equ:F3}, is given at the start of this page. 
Thereafter, we separate the terms on $\beta$ and $\hat{n}$ in $\mathrm{T}^{\mathcal{G}}_{1}$ of~\eqref{Equ:G2} and then simplify it using~\cite[Eq (0.231)]{IntegralBook}. In doing so, we arrive at~\eqref{Equ:G3}. 
As for $\mathrm{T}^{\mathcal{G}}_{2}$ of~\eqref{Equ:G2}, it can be simplified using~\cite[Eq (0.231)]{IntegralBook} to~\eqref{Equ:G4}. 

We next simplify $\mathcal{\tilde{F}}(l,\bar{l},l_i,k_i)$  given in~\eqref{Equ:F}. To this end, similar to that in $\mathcal{\tilde{G}}(k,\bar{k},k_i)$, we first  separate the terms on $s$, $m$, and $m$ in $\mathcal{\tilde{F}}(l,\bar{l},l_i,k_i)$ to obtain~(\ref{Equ:F3}a), where
$\zeta_1=\zeta_1(l,s)=e^{-j\frac{2 \pi}{M}(\frac{s}{\mu}-l)}$, $\zeta_2=\zeta_2(\bar{l},l_i,s)=e^{-j\frac{2 \pi }{M}(\frac{-s+l_i}{\mu}-\bar{l})}$, $\zeta_3=\zeta_3(l_i)=e^{-j\frac{2 \pi (M'-M)l_i }{M'}}$, $\zeta_4=\zeta_4(s)=e^{-j\frac{2 \pi (M'-M)s }{M'}}$, and $\zeta_5=\zeta_5(l,\bar{l},l_i)=e^{j\frac{2 \pi}{M}(l-\bar{l}-\frac{l_i}{\mu})}$.
We then use~\cite[Eq (0.231)]{IntegralBook} in~(\ref{Equ:F3}a) to arrive at~(\ref{Equ:F3}b). 
Finally, substituting~\eqref{Equ:G3},~\eqref{Equ:G4}, and~(\ref{Equ:F3}b) in~\eqref{Equ:ylk2} and then considering $-q=k-\bar{k}-k_i$ and $-\ell=l-\bar{l}-\frac{l_i}{\mu}$ in it, we arrive at the IOR given in Theorem~\ref{Thr:IORCPOTFS-VCP-ECU}, which completes the proof. 
\hfill $\blacksquare$


\end{document}